\documentclass[%
 reprint,
 amsmath,amssymb,
 aps,
 floatfix,
]{revtex4-2}
\usepackage{graphicx}% Include figure files
\usepackage{dcolumn}% Align table columns on decimal point
\usepackage{bm}% bold math
\DeclareMathOperator{\sech}{sech}
\newcommand{\bseq}{\begin{subequations}}
\newcommand{\eseq}{\end{subequations}}
\newcommand{\beq}{\begin{equation}}
\newcommand{\eeq}{\end{equation}}
\newcommand{\beqa}{\begin{eqnarray}}
\newcommand{\eeqa}{\end{eqnarray}}

\begin{document}

\preprint{APS/123-QED}

\title{Mathematical Modeling of Microscale Biology:\\
Ion Pairing, Dielectric
Decrement, and Born Energy in Glycosaminoglycan Brushes}% Force line breaks with \\

\author{Wiliam J Ceely}
 \email{william.ceely@cgu.edu}
\author{Marina Chugunova}%
 \email{marina.chugunova@cgu.edu}
\author{Ali Nadim}%
 \email{ali.nadim@cgu.edu}
\affiliation{%
 Institute of Mathematical Sciences\\
 Claremont Graduate University
}%

\author{James D Sterling}
 \email{jim\_sterling@kgi.edu}
\affiliation{
 Henry E. Riggs School of Applied Life Sciences\\
 Keck Graduate Institute
}%

\date{\today}% It is always \today, today,

\begin{abstract}
Biological macromolecules including nucleic acids, proteins, and glycosaminoglycans are typically anionic and can span domains of up to hundreds of nanometers and even micron length scales. The structures exist in crowded environments that are dominated by weak multivalent electrostatic interactions that can be modeled using mean field continuum approaches that represent underlying molecular nanoscale biophysics. We develop such models for glycosaminoglycan brushes using both steady state modified Poisson-Boltzmann models and transient Poisson-Nernst-Planck models that incorporate important ion-specific (Hofmeister) effects. The results quantify how electroneutrality is attained through ion electrophoresis, dielectric decrement hydration forces, and ion-specific pairing. Brush-Salt interfacial profiles of the electrostatic potential as well as bound and unbound ions are characterized for imposed jump conditions across the interface. The models should be applicable to many intrinsically-disordered biophysical environments and are anticipated to provide insight into the design and development of therapeutics and drug-delivery vehicles to improve human health.
%\begin{description}
%use the optional argument of the \verb+\item+ command to give the category %of each item. 
%\end{description}
\end{abstract}

\maketitle

%\tableofcontents

\section{Introduction}\label{intro}
Many biological structures are characterized by beds of intrinsically-disordered anionic biopolymers. In particular, nucleic acids, proteins, and extracellular glycosaminoglycans (GAGs) are polyelectrolytes that perform functions controlled by their hydration and their neutralization by cations and cationic residues of associated proteins. These anionic beds of slowly-diffusing macromolecules can be considered \textit{fixed} in space over times scales of counterion neutralization phenomena. Although transient phenomena occur with molecular conformation changes, we can place the reference frame at the center of mass of the bed of the macromolecules to make analysis tractable. If the anionic bed is tethered to a tissue- or cell-surface, or a biopolymer is grafted to a surface, simplified structural models known as \textit{brushes} can be defined and there is a long history of brush research in polymer and surface science \cite{de1987polymers, ballauff2006polyelectrolyte,  yu2017multivalent, zimmermann2022quantitative}. 

Another set of simplified structural models of anionic biopolymer beds are spherical \textit{biomolecular condensates}. Here, non-tethered anionic macromolecules interact with counterion atoms and molecules to form two co-existing liquid phases where a dense phase appears in the form of microscale spheres within a dilute phase. The discovery of biomolecular condensates (or \textit{membraneless organelles}) has revolutionized our perspective of biological structure and function, as the formation of condensates helps explain the acceleration of biochemical reaction rates and epigenetic control of biological processes that occur in both intraceullar and extracellular domains \cite{shimobayashi2021nucleation, xue2022phase}. In polymer science, such condensates fall within the broader category of \textit{coacervates} as described in the two recent review articles \cite{astoricchio2020wide, Sing2020recent}.

Pathologies associated with cell surfaces, mucosal surfaces, and membraneless organelles can be recapitulated in laboratory studies of biological structures that can be approximated as brushes and biocondensates. Thus, in a relatively new approach to drug discovery, biomacromolecules can be designed and applied to control microscale biology for therapeutic benefit. Examples include well-known polysaccharides such as heparin and hyaluronic acid as well as more recent cationic-lipids used in mRNA vaccines \cite{granados2021engineering}, poly(acetyl,argyl) glucosamine (PAAG) for mucosal disorders \cite{narayanaswamy2018novel}, as well as cationic arginine-rich peptides for drug delivery applications \cite{edwards2020neuroprotective}.    

In this paper, we develop a broadly-applicable mean-field mathematical model of anionic beds of macromolecules neutralized by cations focusing specifically on \textit{GAG brushes} as a test case. The model aims to elucidate complexities of the biophysics of these molecularly-crowded environments where there is a delicate balance of weak multivalent electrostatic ion-pairing accompanied by release of water and ions upon binding. An ordinary differential equation (ODE) model is developed that depicts steady state electrostatic interactions of a GAG brush in a bulk salt solution taking into account spatial variation of the permittivity (\textit{dielectric decrement}), which leads to
ion Born Hydration energy gradients, and ion pairing between the salt and the GAG brush. This
model is known as a modified Poisson-Boltzmann (MPB) model. We then introduce transient partial differential equations (PDEs) that include the flux and reaction rates of ions that react with counterions to form ion-pairs. The combination of the electrostatics equation and species transport equations are known as Poisson-Nernst-Planck (PNP) equations.

We begin with the derivations of two PB models using different assumptions for the variation of GAG brush-salt permittivity. We then compare our models to a previous model \cite{dean2003molecular}, which does not incorporate dielectric decrement or ion pairing, showing how electrostatic potential and ion concentration profiles match in the limit and we compare
results from both models. We then consider molecular simulation data \cite{sterling2021ion}, which show that dielectric decrement and ion pairing lead to Born energy and a binding energy, respectively. We show how to relate the binding energy to our dissociation constant and compare predictions from our model to the simulation data. We then consider extensions to the model that include two counterions and show that the PNP models are stable and relax to the expected PB steady-states. Finally, we end with a discussion of our model applicability, its shortcomings, and future directions of this work.

\section{The Steady-State Model}\label{deriv_mod}
\begin{figure}
    \centering
    \includegraphics[trim=30 0 30 0, clip, scale=0.5]{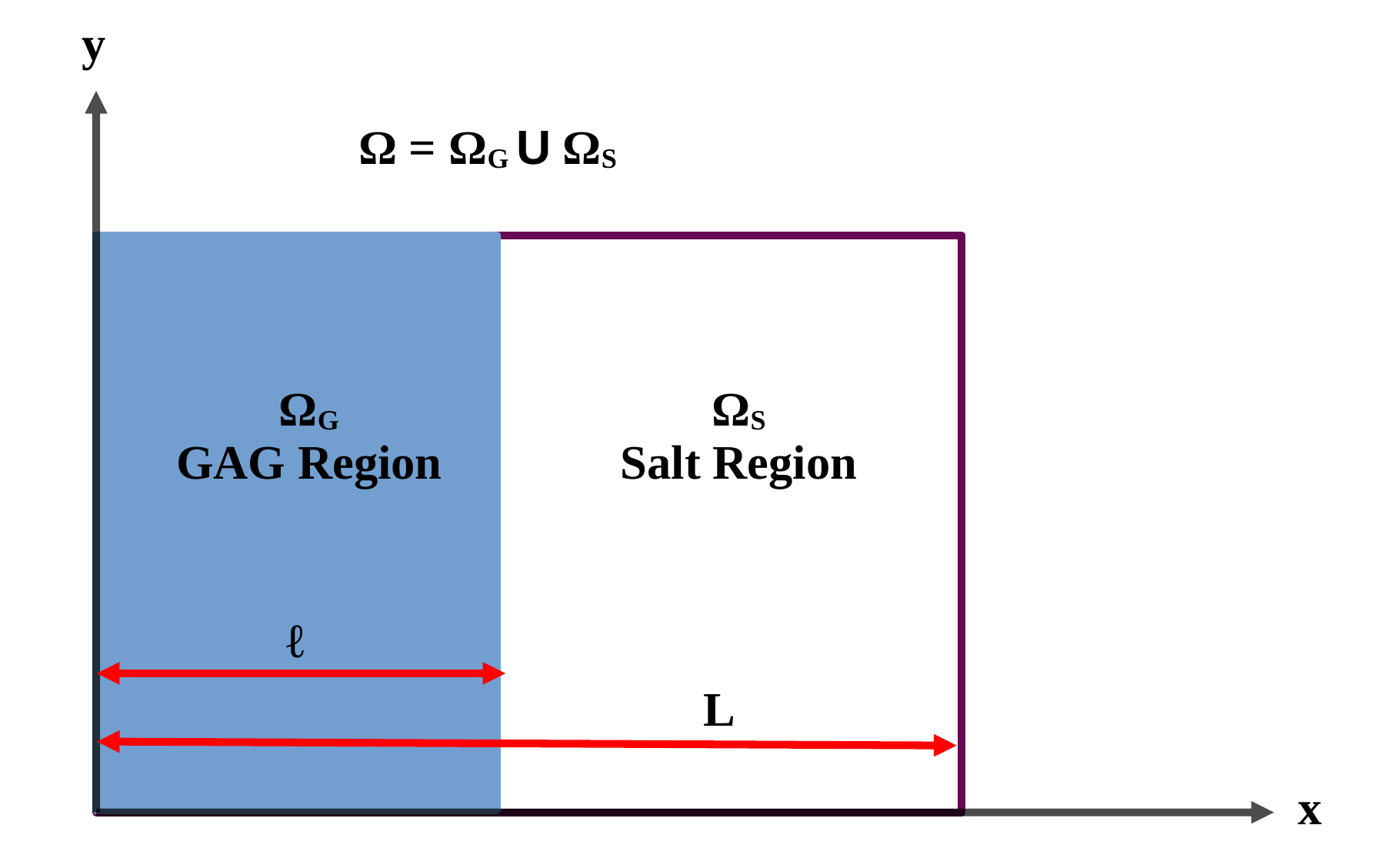}
    \caption{Schematic of model depicting GAG brush region,
    $\Omega_{G} = \{0\leq x \leq \ell\}$, Salt region,
    $\Omega_{S} = \{\ell \leq x \leq L\}$, and total domain,
    $\Omega = \Omega_{G} \cup \Omega_{S}$.}
    \label{mod_fig}
\end{figure}

To derive a mathematical model to describe the electrostatics in Figure~\ref{mod_fig},
we begin, as normal, with the differential form of Gauss's law:
\beq\label{gauss}
\bm{\nabla}\bm{\cdot}\left(\varepsilon\bm{E}\right) = \rho, \qquad \text{in} \ \Omega,
\eeq
where $\bm{E}$ is the electric field, $\rho$ is the total charge density,
and $\varepsilon$ is the permittivity of the medium. Assuming electrostatic conditions
or that the magnetic field variation with time is negligible, the curl of the electric
field can be assumed zero: $\bm{\nabla} \times \bm{E} \approx {\bm 0}$, so that we can define a relationship between the electric field and the potential as:
\beq\label{def_pot}
\bm{E} = -\bm{\nabla}\phi, \qquad \text{in} \ \Omega.
\eeq
Substituting (\ref{def_pot}) into (\ref{gauss}), we arrive at:
\beq
-\bm{\nabla}\bm{\cdot}\left(\varepsilon\bm{\nabla}\phi\right) = \rho, \qquad \text{in} \ \Omega.
\eeq
The total charge density is related to the unbound concentration of the ions present:
\beq
\rho = \sum_{i} z_{i}N_{A}e\left[C_{i}\right],
\eeq
where $z_{i}$ is the valence and $\left[C_{i}\right]$ is the unbound concentration of
ion $i$, respectively, $N_{A}$ is Avogadro's constant, and $e$ is the elementary charge.
We make the simplifying assumption that the variation in potential and ion concentrations
is only present in the $x$-direction, which reduces our model to a single 2nd order ODE.
\beq
-\frac{d}{dx}\left(\varepsilon\frac{d\phi}{dx}\right) = \sum_{i} z_{i}N_{A}e\left[C_{i}\right],
\qquad \text{in} \ \Omega.
\eeq
Finally, we focus on a negatively charged GAG in a monovalent salt solution and thus
$z_{i} = \pm 1$.
\beq\label{mono-salt}
-\frac{d}{dx}\left(\varepsilon\frac{d\phi}{dx}\right) 
= N_{A}e\left(\left[C^{+}\right] - \left[A^{-}\right] - \left[G^{-}\right]\right),
\qquad \text{in} \ \Omega,
\eeq
where $\left[C^{+}\right]$ represents the cation concentration (such as $\text{Na}^{+}$ or $\text{K}^{+}$),
$\left[A^{-}\right]$ represents the anion concentration (such as $\text{Cl}^{-}$), and
$\left[G^{-}\right]$ represents the unbound fixed GAG concentration.

Previous models \cite{dean2003molecular} assume a constant permittivity throughout
the whole domain. However, molecular simulations \cite{sterling2021ion} show a decrease
in the permittivity (dielectric decrement)  within the GAG region. In what follows,
we continue to develop our model using two different assumptions about how the
permittivity and total concentration of the GAG vary throughout the domain. To
this end, we incorporate a Born energy term \cite{bornenergy} to account for this
varying permittivity. The Born energy is given by
\beq
U_{i} = \frac{z_{i}^{2}e^{2}}{8\pi\varepsilon r_{i}},
\eeq
where $r_{i}$ is the Born radius of ion $i$. This is an effective radius and not
a physically measured ion radius.

We also take into account ion pairing (the ability for the cation to bind to the
GAG ions). In a reversible reaction, at equilibrium we must have $k_{1}\left[G^{-}\right]\left[C^{+}\right] = k_{-1}\left[GC\right]$, which leads to:
% \bseq
% \beq
% k_{1}\left[G^{-}\right]\left[C^{+}\right] = k_{-1}\left[GC\right]
% \eeq
\beq\label{ion-pair}
%\implies 
\left[GC\right] = \frac{\left[C^{+}\right]}{(k_{-1}/k_{1})}\left[G^{-}\right] = \frac{\left[C^{+}\right]}{K_{1}}\left[G^{-}\right].
\eeq
%\eseq
Here $\left[GC\right]$ is the concentration of the GAG and the cation that are
bound together, $k_{1}$ is the forward reaction rate constant at which the GAG
and the cation bind together, $k_{-1}$ is the backward reaction rate constant at
which the bound GAG and cation break apart, and
\beq
K_{1} \equiv \frac{k_{-1}}{k_{1}},
\eeq
is known as the dissociation constant and has units of concentration.

The total GAG concentration, $\left[G^{-}\right]_{0}$, is the sum of the unbound
GAG concentration, $\left[G^{-}\right]$, and the concentration of GAG that is
bound to the cation, $\left[GC\right]$. Therefore, $\left[G^{-}\right]_{0} = \left[G^{-}\right] + \left[GC\right]
= \left[G^{-}\right]\left(1+{\left[C^{+}\right]}/{K_{1}}\right)$, resulting in:
% \bseq
% \beqa
% \left[G^{-}\right]_{0} &=& \left[G^{-}\right] + \left[GC\right]
% = \left[G^{-}\right]\left(1+\frac{\left[C^{+}\right]}{K_{1}}\right)
% \eeqa
\beq\label{unbound-gag-conc}
%\implies 
\left[G^{-}\right] = \frac{\left[G^{-}\right]_{0}}
{\left(1+\left[C^{+}\right]/{K_{1}}\right)} \ .
\eeq
%\eseq

\subsection{Piecewise Constant Permittivity and Total GAG Concentration}\label{pw_const}
The simplest assumption that we can make for the varying permittivity and total GAG
concentration is to assume piecewise constant values for both. This allows us to
break the problem into two separate domains where the permittivity and the total GAG
concentration are both constant.

Let 
\beq
\varepsilon = \varepsilon_{0}\varepsilon_{r}\varepsilon_{x},
\eeq
where $\varepsilon_{0}$ is the vacuum permittivity, $\varepsilon_{r}$ is the dielectric
constant of a reference medium, and
\beq
\varepsilon_{x} =
\begin{cases}
    \varepsilon_{S}, \qquad \text{in} \ \Omega_{S}\\
    \varepsilon_{G}, \qquad \text{in} \ \Omega_{G}
\end{cases},
\eeq
are constant scaling factors to the permittivity in the Salt and GAG regions.

In the Salt region, the total GAG concentration is taken to be zero. Thus, we
can represent the ODE of equation (7) as:
\beq\label{salt_ode1}
-\varepsilon_{0}\varepsilon_{r}\varepsilon_{S}\frac{d^{2}\phi}{dx^{2}} = N_{A}e
\left(\left[C^{+}\right] - \left[A^{-}\right]\right) \qquad \text{in} \ \Omega_{S}.
\eeq
We scale all concentrations with some concentration, $C_{0}$. Possible choices for
$C_{0}$ are the bulk concentration of the salt or the total GAG concentration.
Define the dimensionless concentrations as
\beq\label{ndim_con}
c \equiv \frac{\left[C^{+}\right]}{C_{0}}, \quad a \equiv \frac{\left[A^{-}\right]}{C_{0}}.
\eeq
Scale the electric potential, $\phi$, with the thermal voltage
\beq\label{therm_volt}
\frac{k_{B}T}{e} = \frac{N_{A}k_{B}T}{N_{A}e} = \frac{RT}{F},
\eeq
where $k_{B}$, $T$, $R$, and $F$ are Boltzmann constant, absolute temperature, gas constant,
and Faraday constant, respectively. Define the dimensionless potential as
\beq\label{ndim_pot}
y \equiv \frac{\phi}{\left(RT/F\right)}.
\eeq
Substituting (\ref{ndim_con}), (\ref{therm_volt}), and (\ref{ndim_pot}) into
(\ref{salt_ode1}) and dividing both sides by $F$ yields:
\beq
-\frac{\varepsilon_{0}\varepsilon_{r}\varepsilon_{S}RT}{F^{2}C_{0}}\frac{d^{2}y}{dx^{2}}
= c - a.
\eeq
To nondimensionalize lengths, we define a modified Debye length, $\lambda_{D}$ as
\beq\label{debye_def}
\lambda_{D}^{2} = \frac{\varepsilon_{0}\varepsilon_{r}RT}{F^{2}C_{0}},
\eeq
and scale $x$, $\ell$, and $L$ by $\lambda_{D}$. That is,
\beq\label{ndim_len}
\hat{x} = \frac{x}{\lambda_{D}}, \quad \hat{\ell} = \frac{\ell}{\lambda_{D}}, \quad \hat{L} = \frac{L}{\lambda_{D}}.
\eeq
Our dimensionless ODE in the salt region becomes:
\beq
-\frac{d^{2}y}{d\hat{x}^{2}} = \frac{1}{\varepsilon_{S}}\left(c-a\right) \qquad
\text{in} \ \Omega_{S}.
\eeq
In crowded macromolecular environments, details of hydration can substantially impact ion motion and partitioning. Thus, assuming that ions partition according to Boltzmann distributions that combine electrostatic energy with Born hydration energy \cite{lopez2018multiionic},
we can write:
\begin{align}
c &=  \bar{c}\exp\left(-y -{\hat{u}_{c}}/{\varepsilon_{S}}\right),\\
a &=  \bar{a}\exp\left(y - {\hat{u}_{a}}/{\varepsilon_{S}}\right),
\end{align}
where $\hat{u}_{c}={e^{2}}/{8\pi k_{B}T\varepsilon_{0}\varepsilon_{r}r_{c}}$ and $\hat{u}_{a}={e^{2}}/{8\pi k_{B}T\varepsilon_{0}\varepsilon_{r}r_{a}}$ are the dimensionless Born energies in the
reference medium with dielectric constant $\varepsilon_{r}$ for the cation and anion,
respectively. The dimensionless PB equation in the salt region thus takes the explicit form:
\beqa
-\frac{d^{2}y}{d\hat{x}^{2}}= \frac{1}{\varepsilon_{S}}
&\left[\bar{c}\exp\left(-y -{\hat{u}_{c}}/{\varepsilon_{S}}\right)
-\bar{a}\exp\left(y -{\hat{u}_{a}}/{\varepsilon_{S}}\right)\right]\nonumber\\
&\text{in} \ \Omega_{S}.
\eeqa
Since the potential is relative, we need to define the zero reference. We choose
to define the reference potential where electroneutrality is locally met. That is, take
${d^{2}y}/{d\hat{x}^{2}} \equiv 0$ where $y=0$ in $\Omega_{S}$, so that:
\beq
 0 = \bar{c}\exp(-{\hat{u}_{c}}/{\varepsilon_{S}})
-\bar{a}\exp(-{\hat{u}_{a}}/{\varepsilon_{S}})
\eeq
This can be written as $\widetilde{c} = \widetilde{a}$, where $\widetilde{c}=\bar{c}\exp(-{\hat{u}_{c}}/{\varepsilon_{S}})$ and $\widetilde{a}=\bar{a}\exp(-{\hat{u}_{a}}/{\varepsilon_{S}})$ are rescaled dimensionless
concentrations in the bulk salt solution.

The final version of the governing equations, known as the Born-energy augmented Poisson-Boltzmann equation \cite{wang2010fluctuation}, reduce to the following. In the Salt region we have:
\beq\label{salt_final}
-\frac{d^{2}y}{d\hat{x}^{2}}= \frac{\widetilde{c}}{\varepsilon_{S}}
\left[e^{-y} - e^{y}\right] \qquad \text{in} \ \Omega_{S}.
\eeq
In the GAG region, the total GAG concentration, $\left[G^{-}\right]_{0}$, is taken
to be constant, whereby:
\beqa
-\varepsilon_{0}\varepsilon_{r}\varepsilon_{G}\frac{d^{2}\phi}{dx^{2}}
% &=& N_{A}e\left(\left[C^{+}\right] - \left[A^{-}\right] - \left[G^{-}\right]\right)\nonumber\\
&=& N_{A}e\left(\left[C^{+}\right] - \left[A^{-}\right] - \frac{\left[G^{-}\right]_{0}}
{\left(1+\left[C^{+}\right]/{K_{1}}\right)}\right)\nonumber\\
& &\text{in} \ \Omega_{G}.
\eeqa
Using the same scaling parameters ($C_{0}$, $RT/F$, $\lambda_{D}$) from the salt region derivation we obtain:
\beq
-\frac{d^{2}y}{d\hat{x}^{2}}= \frac{1}{\varepsilon_{G}}\left[c - a 
-\frac{\bar{g}}{1+c/\widetilde{K}_{1}}\right] \qquad \text{in} \ \Omega_{G},
\eeq
where $\bar{g} = \left[G^{-}\right]_{0}/C_{0}$ and $\widetilde{K}_{1} = K_{1}/C_{0}$.
Again, assuming Boltzmann distributions combined with Born energy for the mobile ions:
\begin{align}
c &= \widetilde{c}\exp\left(-y-\hat{u}_{c}
\left[\frac{1}{\varepsilon_{G}}-\frac{1}{\varepsilon_{S}}\right]\right),\\
a &= \widetilde{c}\exp\left(y-\hat{u}_{a}
\left[\frac{1}{\varepsilon_{G}}-\frac{1}{\varepsilon_{S}}\right]\right).
\end{align}
The final version of the ODE in the GAG region is:
\begin{widetext}
\beqa\label{gag_final}
-\frac{d^{2}y}{d\hat{x}^{2}}=& \frac{\widetilde{c}}{\varepsilon_{G}}
\left\{\exp\left(-y-\hat{u}_{c}
\left[\frac{1}{\varepsilon_{G}}-\frac{1}{\varepsilon_{S}}\right]\right)
-\exp\left(y-\hat{u}_{a}
\left[\frac{1}{\varepsilon_{G}}-\frac{1}{\varepsilon_{S}}\right]\right)\right\}
-\frac{1}{\varepsilon_{G}}\left\{\frac{\bar{g}}{1+\left(\widetilde{c}/\widetilde{K}_{1}\right)
\exp\left(-y-\hat{u}_{c}
\left[\frac{1}{\varepsilon_{G}}-\frac{1}{\varepsilon_{S}}\right]\right)}\right\}\nonumber\\
&\qquad \text{in} \ \Omega_{G}.
\eeqa
\end{widetext}
The value of the dimensionless potential, $y$, that makes the right-hand side zero is known as the Donnan potential, $y_{D}$.

Thus, the model assuming piecewise constant permittivity and total GAG concentration
is complete with equations (\ref{salt_final}) and (\ref{gag_final}) along with
the boundary conditions:
\bseq\label{bc-piescewise}
\beq\label{bc1-piecewise}
y(\hat{\ell}^{-}) = y(\hat{\ell}^{+}),
\eeq
\beq
\varepsilon_{G}\frac{d}{d\hat{x}}y(\hat{\ell}^{-}) = \varepsilon_{S}\frac{d}{d\hat{x}}y(\hat{\ell}^{+}).
\eeq
\beq
-\frac{d}{d\hat{x}}y(0) = \hat{\sigma}_{2}.
\eeq
\beq\label{bc4-piecewise}
\frac{d}{d\hat{x}}y(\hat{L}) = \hat{\sigma}_{1}.
\eeq
\eseq
The first two conditions equate the potential and positive and negative surface charge densities at the interface. 
For electroneutrality to hold overall, we must have $\hat{\sigma}_{1}=-\hat{\sigma}_{2}$. These are the 
surface charge densities at the two boundaries. 
\bigskip

\subsection{Smoothly Varying Permittivity and Total GAG Concentration}\label{smooth}
Perhaps a more accurate assumption for the varying permittivity and total GAG
concentration is to take them to vary smoothly with respect to $x$ rather than being discontinuous at the interface. The same PB equation thus holds throughout the domain:
\beq
-\varepsilon_{0}\varepsilon_{r}\frac{d}{dx}\left(\varepsilon_{1}(x)\frac{d\phi}{dx}\right)
= N_{A}e\left(\left[C^{+}\right] - \left[A^{-}\right] - \left[G^{-}\right]\right).
\eeq
Using the same scaling parameters ($C_{0}$, $RT/F$, $\lambda_{D}$, $\hat{u}_{c}$,
and $\hat{u}_{a}$) from section \ref{pw_const}, the dimensionless ODE model $-\left[\varepsilon_{1}(\hat{x}) y'(\hat{x})\right]' = c - a - g$ becomes:
\begin{widetext}
\beq\label{smooth_final}
-\varepsilon_{1}(\hat{x})\frac{d^{2}y}{d\hat{x}^{2}}
-\frac{d \varepsilon_{1}(\hat{x})}{d\hat{x}}\frac{dy}{d\hat{x}}
= \bar{c}\exp\left(-y - \frac{\hat{u}_{c}}{\varepsilon_{1}(\hat{x})}\right)
- \bar{a}\exp\left(y - \frac{\hat{u}_{a}}{\varepsilon_{1}(\hat{x})}\right)
- \frac{\bar{g}(\hat{x})}{1+\bar{c}/\widetilde{K}_{1}
\exp\left(-y - {\hat{u}_{c}}/{\varepsilon_{1}(\hat{x})}\right)} \qquad \text{in} \ \Omega,
\eeq
\end{widetext}
with boundary conditions:
\bseq\label{bc-smooth}
\beq\label{bc1-smooth}
-\frac{d}{d\hat{x}}y(0) = \hat{\sigma}_{2},
\eeq
\beq\label{bc2-smooth}
\frac{d}{d\hat{x}}y(\hat{L}) = \hat{\sigma}_{1}.
\eeq
\eseq
To complete the model, we need to define the smooth functions $\varepsilon_{1}(\hat{x})$
and $\bar{g}(\hat{x})$. A convenient choice for a smooth function is the hyperbolic
tangent. Let
\beq
f(\hat{x}) = \frac{1}{2}
\left[\tanh\left(\frac{1-{\hat{x}}/{\hat{\ell}}}{\alpha}\right)+1\right],
\eeq
which has the first derivative
\beq
f'(\hat{x}) = -\frac{1}{2\hat{\ell}\alpha}\left[\sech\left(
\frac{1-{\hat{x}}/{\hat{\ell}}}{\alpha}\right)+1\right]^{2}.
\eeq
The transition is centered at $\hat{x}=\hat{\ell}$, and $\alpha > 0$ controls the
transition length. We then define
\beq
\bar{g}(\hat{x}) \equiv g_{0}f(\hat{x}),
\eeq
\beq
\varepsilon_{1}(\hat{x}) \equiv (\varepsilon_{G}-\varepsilon_{S})f(\hat{x}) + \varepsilon_{S},
\eeq
\beq
\frac{d\varepsilon_{1}(\hat{x})}{dx} \equiv (\varepsilon_{G}-\varepsilon_{S})f'(\hat{x}).
\eeq
We can express $\bar{a}$ in terms of $\bar{c}$ by defining the first and second
derivative of $y$ to be zero when $y=0$, noting that $\varepsilon_{1} \approx \varepsilon_{S}$
and $\bar{g} \approx 0$ in this case. With this we see that
\beq
\bar{a} = \bar{c}\exp\left(\frac{\hat{u}_{a}-\hat{u}_{c}}{\varepsilon_{S}}\right).
\eeq
We argue that this model is equivalent to the model in \S \ref{pw_const} in the limit.
With the proposed smooth function, for $\alpha \ll 1$, the permittivity and total
GAG concentration are essentially piecewise constant except for near the interface
$(\hat{x} \approx \hat{\ell})$. So the model in \S \ref{pw_const} is an approximation
to this model away from the interface. Further, we also note that as $\alpha \to 0$,
\beq f(\hat{x})
\to \begin{cases}
1, \qquad \text{in} \ \Omega_{G}\\
0, \qquad \text{in} \ \Omega_{S}
\end{cases},
\eeq
that is, $f(\hat{x})$ approaches a piecewise constant function, and this version
of the model approaches that of section \ref{pw_const}.

\section{Previous Volume Charge Model: GAG Brush Near a Charged Surface}\label{prev_mod}

\subsection{Brief Description of the Model}\label{mit_mod}
Dean et al.~\cite{dean2003molecular} provide three mathematical PB models to describe
the electrostatic interactions of a negatively charged chondroitin sulfate
GAG in a bulk NaCl salt solution. The basic form of the models is
\beq
\nabla^{2}\Phi = \frac{2FC_{0}}{\varepsilon_{w}}\sinh\left(\frac{F\Phi}{RT}\right)
- \frac{\rho_{\text{fix}}}{\varepsilon_{w}},
\eeq
where $C_{0}$ is the bath concentration of NaCl, $\rho_{\text{fix}}$ is a fixed
charge density term and $\varepsilon_{w}$ is the permittivity of the bulk solution.

\begin{figure}
    \centering
    \includegraphics[trim=30 0 20 0, clip, scale=0.65]{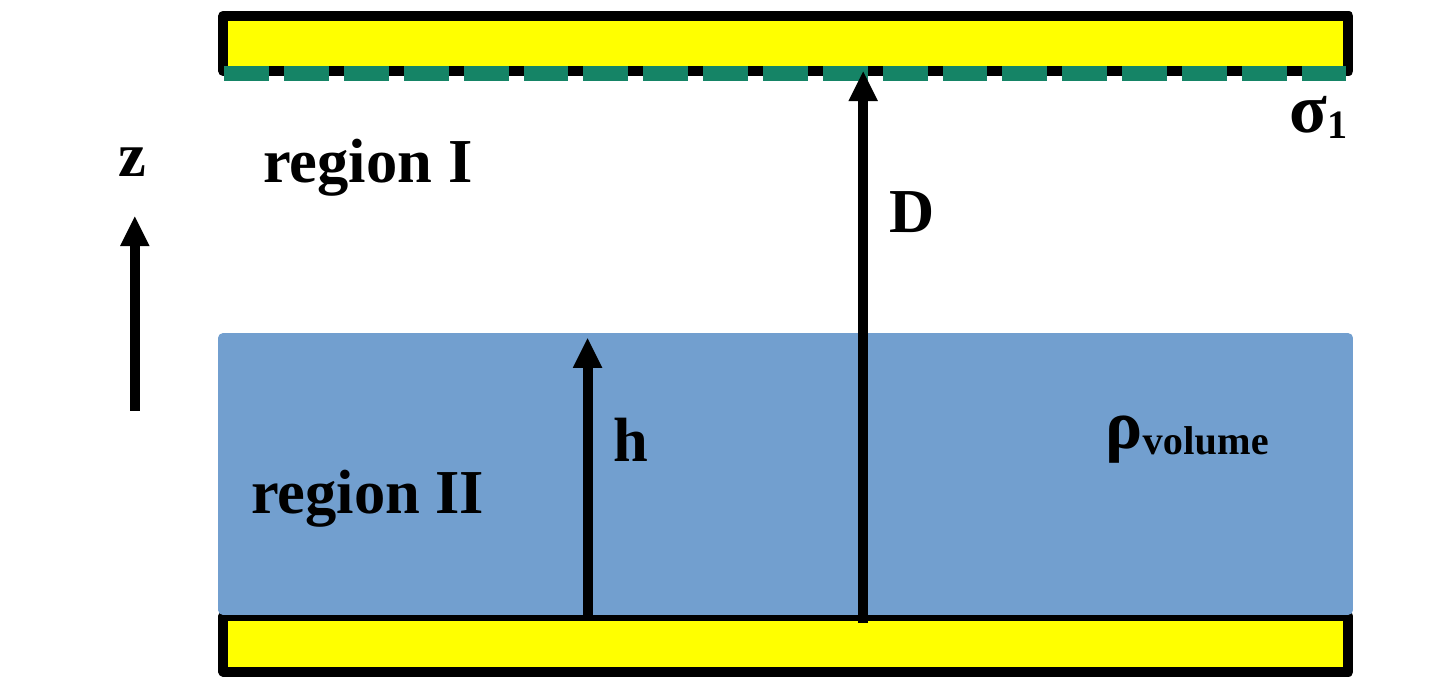}
    \caption{Schematic of constant volume charge model depicting GAG brush with
    constant volume charge density, $\rho_{\text{volume}}$, in region II of height,
    h, bulk salt in region I extending from $h \leq z \leq D$, and surface charge
    density, $\sigma_{1}$. Recreated from \cite{dean2003molecular}.}
    \label{mit_mod_fig}
\end{figure}

The volume charge model in Figure~\ref{mit_mod_fig} approximates the GAG brush
as a fixed uniform volume charge density of height, $h$. Inside the GAG brush
region, the ODE has the form
\beq\label{mit_gag_final}
\nabla^{2}\Phi = \frac{2FC_{0}}{\varepsilon_{w}}\sinh\left(\frac{F\Phi}{RT}\right)
- \frac{\rho_{\text{volume}}}{\varepsilon_{w}},
\eeq
and in the salt region, the ODE has the form
\beq\label{mit_salt_final}
\nabla^{2}\Phi = \frac{2FC_{0}}{\varepsilon_{w}}\sinh\left(\frac{F\Phi}{RT}\right).
\eeq
The boundary conditions are given as
\bseq
\beq
\frac{\partial}{\partial z}\Phi(0) = 0,
\eeq
\beq
\frac{\partial}{\partial z}\Phi(D) = \frac{\sigma_{1}}{\varepsilon_{w}},
\eeq
\beq
\Phi(h^{-}) = \Phi(h^{+}),
\eeq
\beq\label{mit_last_eqn}
\frac{\partial}{\partial z}\Phi(h^{-}) = \frac{\partial}{\partial z}\Phi(h^{+}).
\eeq
\eseq

\subsection{Relationship Between Models}\label{rel_mit_mod}
The models in \cite{dean2003molecular} assume a constant permittivity throughout
both the salt and GAG regions. In our model, this would imply $\varepsilon_{S}=\varepsilon_{G}=1$
and thus $\varepsilon_{w} = \varepsilon_{0}\varepsilon_{r}$. Nondimensionalizing the ODEs,
using the same scaling parameters from our model, and replacing $z$ with $x$:
\beq
\frac{d^{2}y}{d\hat{x}^{2}} = 2\bar{c}\sinh\left(y\right) = \bar{c}\left[e^{y} - e^{-y}\right] \quad \mbox{in I},
\eeq
\beq
\frac{d^{2}y}{d\hat{x}^{2}} = \bar{c}\left[e^{-y} - e^{-y}\right]
-\frac{\rho_{\text{volume}}}{FC_{0}} \quad \mbox{in II}.
\eeq
% \beqa
% \frac{d^{2}y}{d\hat{x}^{2}} &=& 2\bar{c}\sinh\left(y\right)\nonumber\\
% \to -\frac{d^{2}y}{d\hat{x}^{2}} &=& \bar{c}\left[e^{-y} - e^{y}\right],
% \eeqa
% \beqa
% \frac{d^{2}y}{d\hat{x}^{2}} &=& 2\bar{c}\sinh\left(y\right)
% -\frac{\rho_{\text{volume}}}{FC_{0}}\nonumber\\
% \to -\frac{d^{2}y}{d\hat{x}^{2}} &=& \bar{c}\left[e^{-y} - e^{y}\right]
% +\frac{\rho_{\text{volume}}}{FC_{0}}.
% \eeqa
These are similar to our model in equations (\ref{salt_final}) and (\ref{gag_final}),
with two important differences. The first is that \cite{dean2003molecular} does
not include a Born energy term. However, because the permittivity is assumed constant,
the Born energy would also be constant. Therefore, it could be lumped into the
dimensionless concentration $\bar{c}$, which matches our $\widetilde{c}$. The
second difference is that \cite{dean2003molecular} ignores any ion pairing. In
other words, the total GAG concentration is equal to the unbound GAG concentration.
If we allow the dissociation constant to approach $\infty$, then 
$\left[G^{-}\right] \to -\rho_{\text{volume}}$, and we see that our model
approaches the model in \cite{dean2003molecular} in the limit.

Table~\ref{mit_params} summarizes the parameters used in \cite{dean2003molecular}.
Table~\ref{mit_input_params} summarizes the input parameters needed for our model
described in Section \ref{pw_const} and how they relate to the values in
Table~\ref{mit_params}. Not that since the permittivity is constant,
the terms $\hat{u}_{c}\left[1/{\varepsilon_{G}}-1/{\varepsilon_{S}}\right]$
and $\hat{u}_{a}\left[{1}/{\varepsilon_{G}}-{1}/{\varepsilon_{S}}\right]$ from
equation (\ref{gag_final}) are zero and not needed.

\begin{table}
  \caption{Parameters from \cite{dean2003molecular}.}
  \label{mit_params}
  \begin{ruledtabular}
  \begin{tabular}{ll}
    \textbf{Parameters}                       & \textbf{Volume Model}   \\
    \hline
    $\sigma_{1}$ (C/m$^{2}$)                  & $-0.015$                \\
    $\sigma_{2}$ (C/m$^{2}$)                  & $0$                \\
    $Q$ (C)                                   & $-8.00 \times 10^{-18}$ \\
    $s$ (nm)                                  & $6.5$                   \\
    $h$ (nm)                                  & $20$                    \\
    $D$ (nm)                                  & $30$                    \\
    $\rho_{\text{volume}}$ (C/m$^{3}$)        & $Q/(s^{2}h)$            \\
    $T$ (K)                                   & $298$                   \\
    $\varepsilon_{w}$ (C$^{2}$/N$\cdot$m$^{2}$)  & $6.92 \times 10^{-10}$  \\
    $C_{0}$ (M)                               & $0.01, \ 0.1, \ 1.0$      \\
  \end{tabular}
  \end{ruledtabular}
\end{table}
\begin{table*}
  \caption{Input Parameters to Piecewise Constant Permittivity and
  Total GAG Concentration Model.}
  \label{mit_input_params}
  \begin{ruledtabular}
  \begin{tabular}{lllll}
    \textbf{Parameter} & \textbf{Notes}
    & \textbf{Value for $C_{0}=0.01$} & \textbf{Value for $C_{0}=0.1$} 
    & \textbf{Value for $C_{0}=1$} \\
    \hline
    $\varepsilon_{r}$ & $\varepsilon_{0}/\varepsilon_{w}$ & $78.155$ & $78.155$ & $78.155$ \\
    $\varepsilon_{S}$ & Constant $\varepsilon_{w}$ & $1$ & $1$ & $1$ \\
    $\varepsilon_{G}$ & Constant $\varepsilon_{w}$ & $1$ & $1$ & $1$ \\
    $\hat{\ell}$ & $h/\lambda_{D}$ & $4.660$ & $14.737$ & $46.603$ \\
    $\hat{L}$ & $D/\lambda_{D}$ & $6.990$ & $22.106$ & $69.904$ \\
    $\widetilde{K}_{1}$ & No Binding & $10^{16}$ & $10^{16}$ & $10^{16}$ \\
    $\widetilde{c}$ & $C_{0}/C_{0}$ & $1$ & $1$ & $1$ \\
    $\bar{g}$ & $\dfrac{-\rho_{\text{volume}}}{FC_{0}}$ & $9.812$
    & $9.812\times 10^{-1}$ & $9.812\times 10^{-2}$ \\
    $\hat{\sigma}_{1}$ & $\dfrac{\sigma_{1}F\lambda_{D}}{\varepsilon_{w}RT}$
    & $-3.623$ & $-1.146$ & $-0.362$ \\
    $\hat{\sigma}_{2}$ & $\dfrac{-\sigma_{2}F\lambda_{D}}{\varepsilon_{w}RT}$ & $0$
    & $0$ & $0$ \\
  \end{tabular}
  \end{ruledtabular}
\end{table*}

\subsection{Comparison of Results}\label{comp_mit_mod}
\begin{figure}
    \centering
    (a)\\
    \includegraphics[scale=0.55]{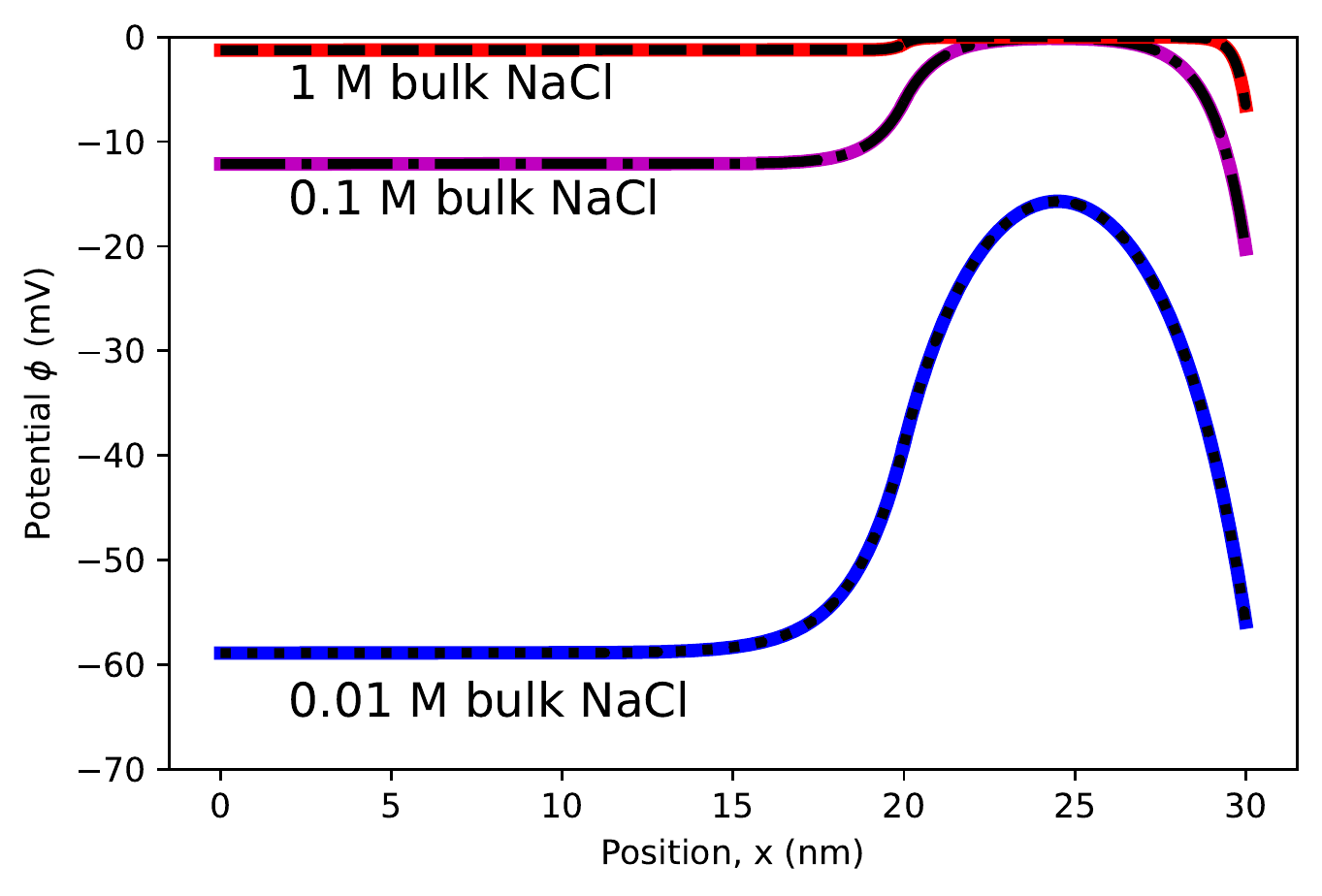}\\
    (b)\\
    \includegraphics[scale=0.55]{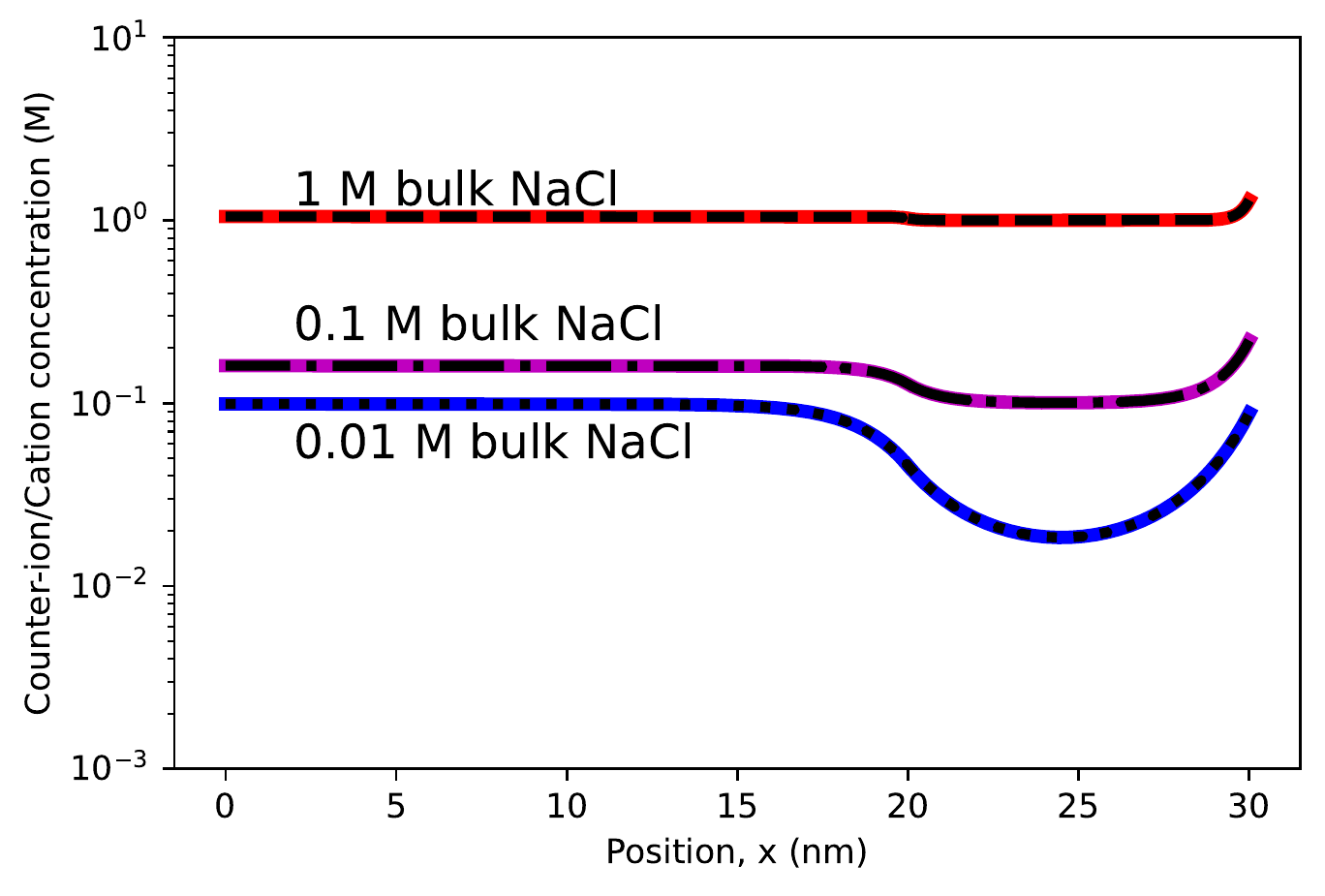}\\
    (c)\\
    \includegraphics[scale=0.55]{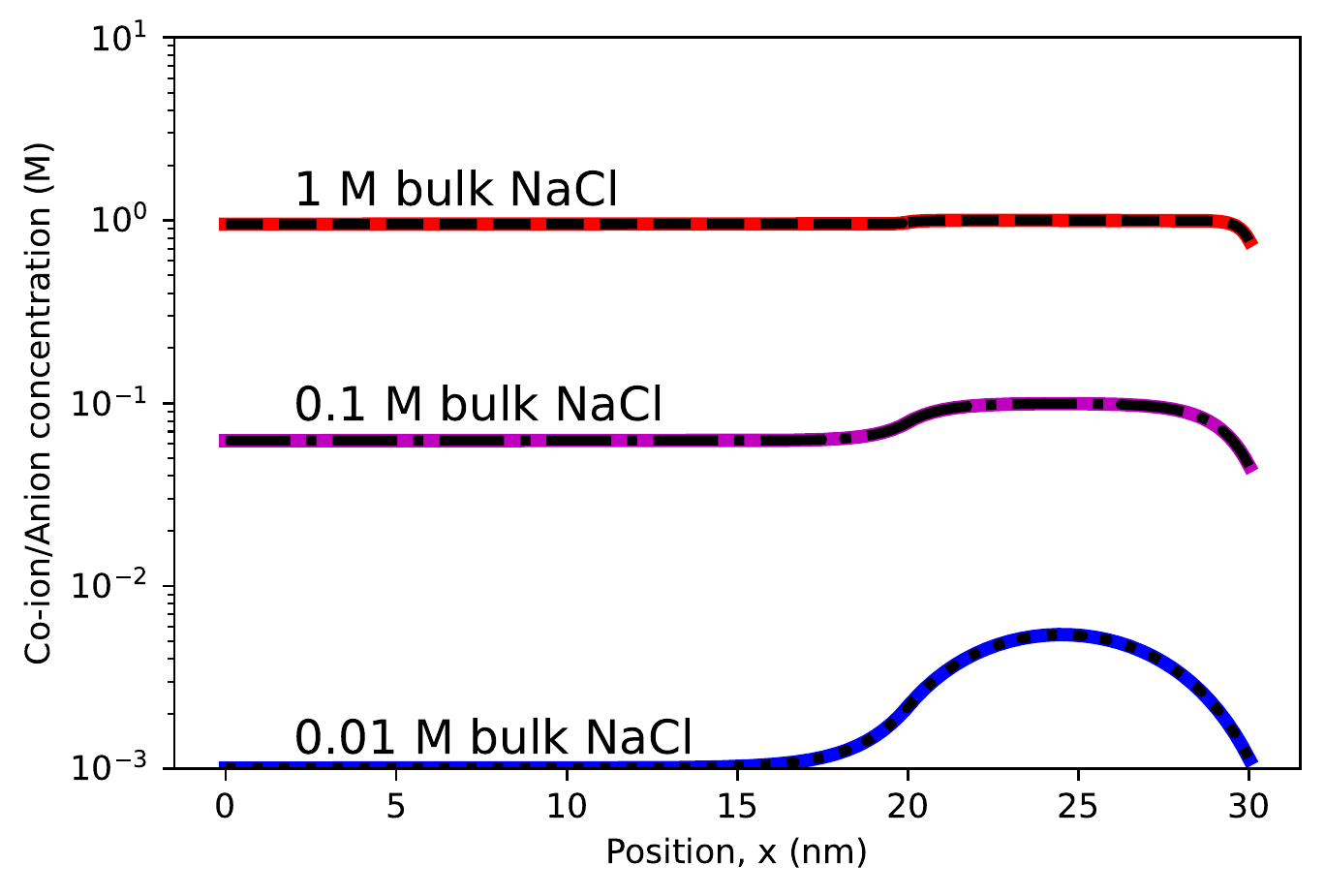}
    \caption{(a) Potential, $\phi$ (mV), (b) cation concentration (M), and (c) anion
    concentration (M) using parameters from Tables \ref{mit_params} and
    \ref{mit_input_params} for $C_{0}$ = 0.01 M, 0.1 M, and 1.0 M. Solid colored
    lines represent the output using the model from \cite{dean2003molecular} and
    dashed/dotted lines represent the output using the model from Section \ref{pw_const}.
    Vertical axis scale chosen to match figures in \cite{dean2003molecular}.}
    \label{mit_out_fig}
\end{figure}

Using the model equations from \cite{dean2003molecular} (summarized in equations
(\ref{mit_salt_final})--(\ref{mit_last_eqn})), with the parameters summarized
in Table~\ref{mit_params}, we replicated the predicted profiles for potential,
$\phi$, and both the cation and anion concentration. We then used
our model from Section \ref{pw_const}, with the parameters summarized in
Table~\ref{mit_input_params} to generate the same profiles. The results are
shown in Figure~\ref{mit_out_fig}, with the solid colored lines representing the
predictions from the model in \cite{dean2003molecular} and the dashed/dotted lines
representing the predictions from the model in Section \ref{pw_const}. As can be
seen, the predictions are in perfect agreement, further showing that in the limit,
our model approaches the volume charge model in \cite{dean2003molecular}.

\section{Previous Molecular Dynamics Simulations: GAG Brush-Salt Interface}\label{prev_sim}

\subsection{Brief Description of Molecular Simulations}\label{sim_data}
Sterling et al.~\cite{sterling2021ion} performed all-atom molecular dynamics
nanoscale simulations of hyaluronic acid and heparin GAG brushes in NaCl and KCl
solutions. The results of the simulation are reported for the average brush values, represented by subscript \textit{b} relative to values in the bulk salt layer denoted by subscript \textit{o} including the Donnan potential, the dielectric decrement (variation of the permittivity), Born energy and the total concentrations of the ions. Thus, a Born-modified Boltzmann partitioning with ion-pair
binding is proposed as:
\beqa
c_{i,b} = c_{i,o}\exp\bigg(-\frac{q_{i}\varphi_{b}}{k_{B}T}
&&-\frac{-\Delta G_{i}^{s}}{k_{B}T}\frac{\varepsilon_{w}}{\varepsilon_{w}-1}\left[\frac{1}{\varepsilon_{b}}
-\frac{1}{\varepsilon_{o}}\right]\nonumber\\&&-\frac{\Delta\mu_{i}}{k_{B}T}\bigg),
\eeqa
where the first term represents the ion charge $q_{i}$ in Donnan potential $\varphi_{b}$,
the second term is the Born energy term, the third term represents the ion-pair
binding free energy between ion atoms and brush anionic charges.

\subsection{Relationship Between Simulation Data and Model}\label{rel_sim_data}
In \cite{sterling2021ion}, the simulation models consisted of a bulk solution
occupying $z = -240 \textup{\AA} \ \text{to} \ 240 \textup{\AA}$ and GAG
brushes of varying lengths, $2\ell$, centered at $z=0$. Our model only considers
the positive half of this domain $x = 0 \ \text{to} \ 240 \textup{\AA}$ with the
GAG brush occupying $x = 0 \ \text{to} \ \ell \textup{~\AA}$. With zero-gradient boundary conditions at 0 and $\ell$, we model the interface between an infinite brush and infinite salt layer. The conditions relating quantities far from the interface represent so-called \textit{jump conditions} that are analogous to the Rankine-Hugoniot conditions that are imposed across shock wave interfaces.

In the results from \cite{sterling2021ion}, total concentrations are used instead
of those of the unbound charged ions, and a cation binding energy term is introduced. Modifying equation
(\ref{gag_final}) to incorporate the binding energy and use total concentrations:
\beqa\label{gag_total}
-\frac{d^{2}y}{d\hat{x}^{2}}=&& \frac{\widetilde{c}}{\varepsilon_{G}}
\bigg\{\exp\left(-y-\hat{u}_{c}
\left[\frac{1}{\varepsilon_{G}}-\frac{1}{\varepsilon_{S}}\right]
-\frac{\Delta\mu}{k_{B}T}\right)\nonumber\\
&&-\exp\left(y-\hat{u}_{a}
\left[\frac{1}{\varepsilon_{G}}-\frac{1}{\varepsilon_{S}}\right]\right)\bigg\}
-\frac{\bar{g}}{\varepsilon_{G}}.
\eeqa
When $y=y_{D}$,
\beqa
\bar{g} = \widetilde{c}\bigg\{&&\exp\left(-y_{D}-\hat{u}_{c}
\left[\frac{1}{\varepsilon_{G}}-\frac{1}{\varepsilon_{S}}\right]
-\frac{\Delta\mu}{k_{B}T}\right)\nonumber\\
&&-\exp\left(y_{D}-\hat{u}_{a}
\left[\frac{1}{\varepsilon_{G}}-\frac{1}{\varepsilon_{S}}\right]\right)\bigg\}.
\eeqa
To find a relationship between the binding energy and the dissociation constant,
substitute this into (\ref{gag_final}) with $y=y_{D}$ and solve for $\widetilde{K_{1}}$:
\begin{widetext}
\beq
\widetilde{K}_{1} = \frac{\widetilde{c}\left\{\exp\left(-y_{D}-\hat{u}_{c}\left[
{1}/{\varepsilon_{G}} - {1}/{\varepsilon_{S}}\right]\right)
-\exp\left(y_{D}-\hat{u}_{a}\left[{1}/{\varepsilon_{G}}
-{1}/{\varepsilon_{S}}\right]\right)\right\}}{\exp\left(-{\Delta\mu}/{k_{B}T}\right)-1}.
\eeq
\end{widetext}
The above relationships are only valid at the Donnan potential, to compute the
binding energy for any potential, set the right-hand sides of (\ref{gag_final})
and (\ref{gag_total}) equal and solve for $\exp\left(-{\Delta\mu}/{(k_{B}T)}\right)$:
\beq
\exp\left(-\frac{\Delta\mu}{k_{B}T}\right) = 1 + \frac{\bar{g}}
{\widetilde{K}_{1}+\widetilde{c}\exp\left(-y-\hat{u}_{c}\left[\frac{1}{\varepsilon_{G}}
-\frac{1}{\varepsilon_{S}}\right]\right)}.
\eeq
Similarly, using equation (\ref{smooth_final}) instead of (\ref{gag_final}), and
noting that $\varepsilon_{1}(\hat{x}) = \varepsilon_{G}$ deep in the brush where the
Donnan potential occurs, we find the relationships to be:
\beq\label{bind_to_K1}
\widetilde{K}_{1} = \frac{\bar{c}\exp\left(-y_{D}-\frac{\hat{u}_{c}}
{\varepsilon_{G}}\right)
-\bar{a}\exp\left(y_{D}-\frac{\hat{u}_{a}}{\varepsilon_{G}}\right)}
{\exp\left(-\frac{\Delta\mu}{k_{B}T}\right)-1},
\eeq
or
\beq
\exp\left(-\frac{\Delta\mu}{k_{B}T}\right) = 1 + \frac{\bar{g}}
{\widetilde{K}_{1}+\bar{c}\exp\left(-y-\frac{\hat{u}_{c}}{\varepsilon_{1}(\hat{x})}\right)}.
\eeq
This can be used to compute the total cation concentration in addition to the unbound concentration.

\subsection{Comparison of Results}\label{comp_sim_data}

Table~\ref{sim_params} summarizes the parameters needed from \cite{sterling2021ion}
to generate the input parameters to the model in Section \ref{smooth}.
The input parameters are listed in Table~\ref{sim_input_params}. Based on the
literature, we expected the Born radius for Chloride to be $2.26$ \textup{\AA}
\cite{sun2020analysis}. However, by using this value in our model, we found that
Chloride would not be excluded from the brush as in the molecular simulation results.
This lead to a negative concentration of unbound GAG ions, a negative dissociation
constant, and/or the inability to match the Donnan potential. By reducing the Born
radius for Chloride by a factor of 10, our model was able to overcome these issues.
Table~\ref{sim_input_params} shows the values of $\hat{u}_{a}$, $\widetilde{K}_{1}$,
and $\bar{a}$ based on both the expected Born radius and the actual value
of $0.226$ \textup{\AA} used in our model.

\begin{figure}
    \centering
    (a)\\
    \includegraphics[scale=0.55]{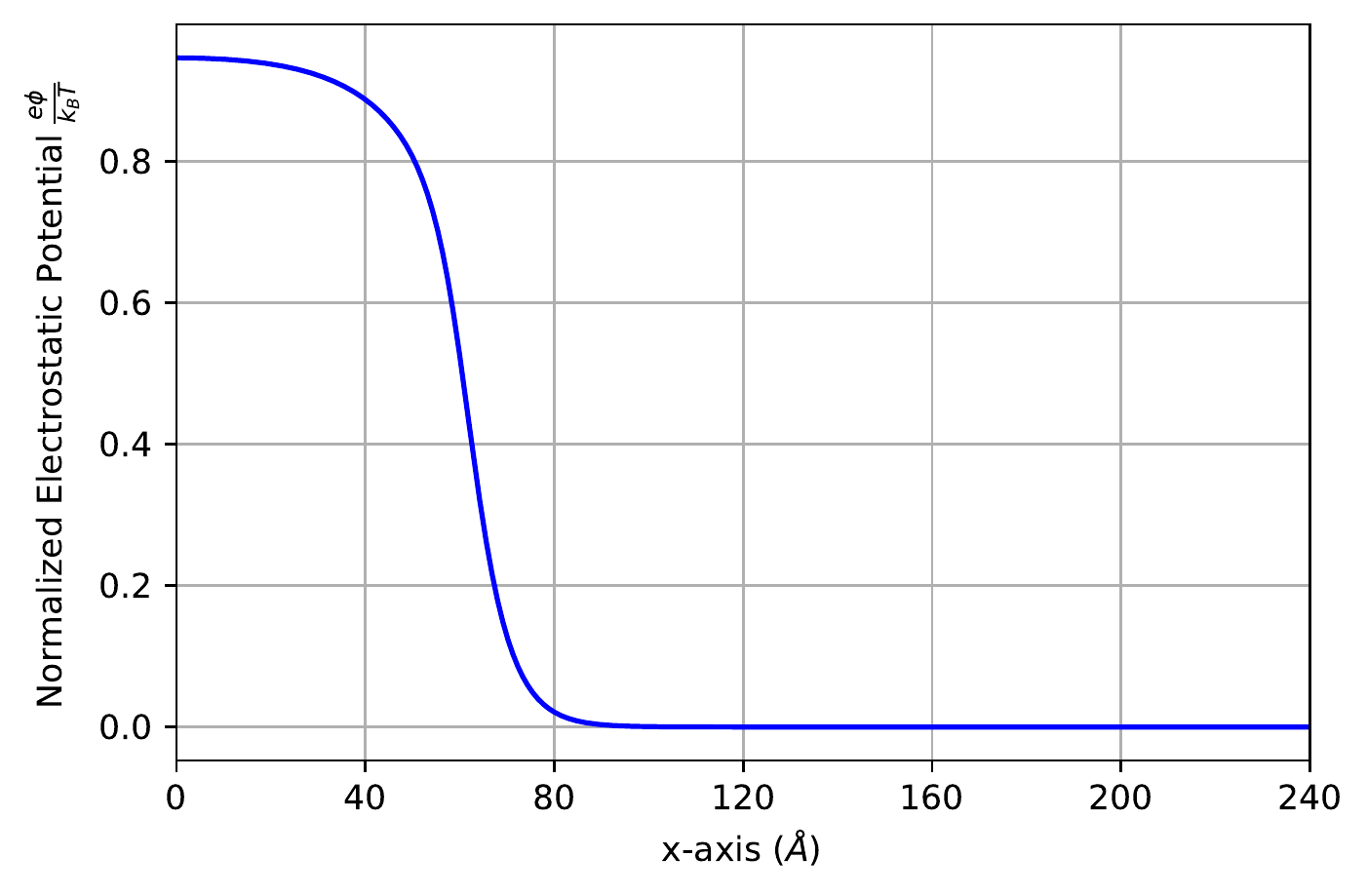}\\
    (b)\\
    \includegraphics[scale=0.55]{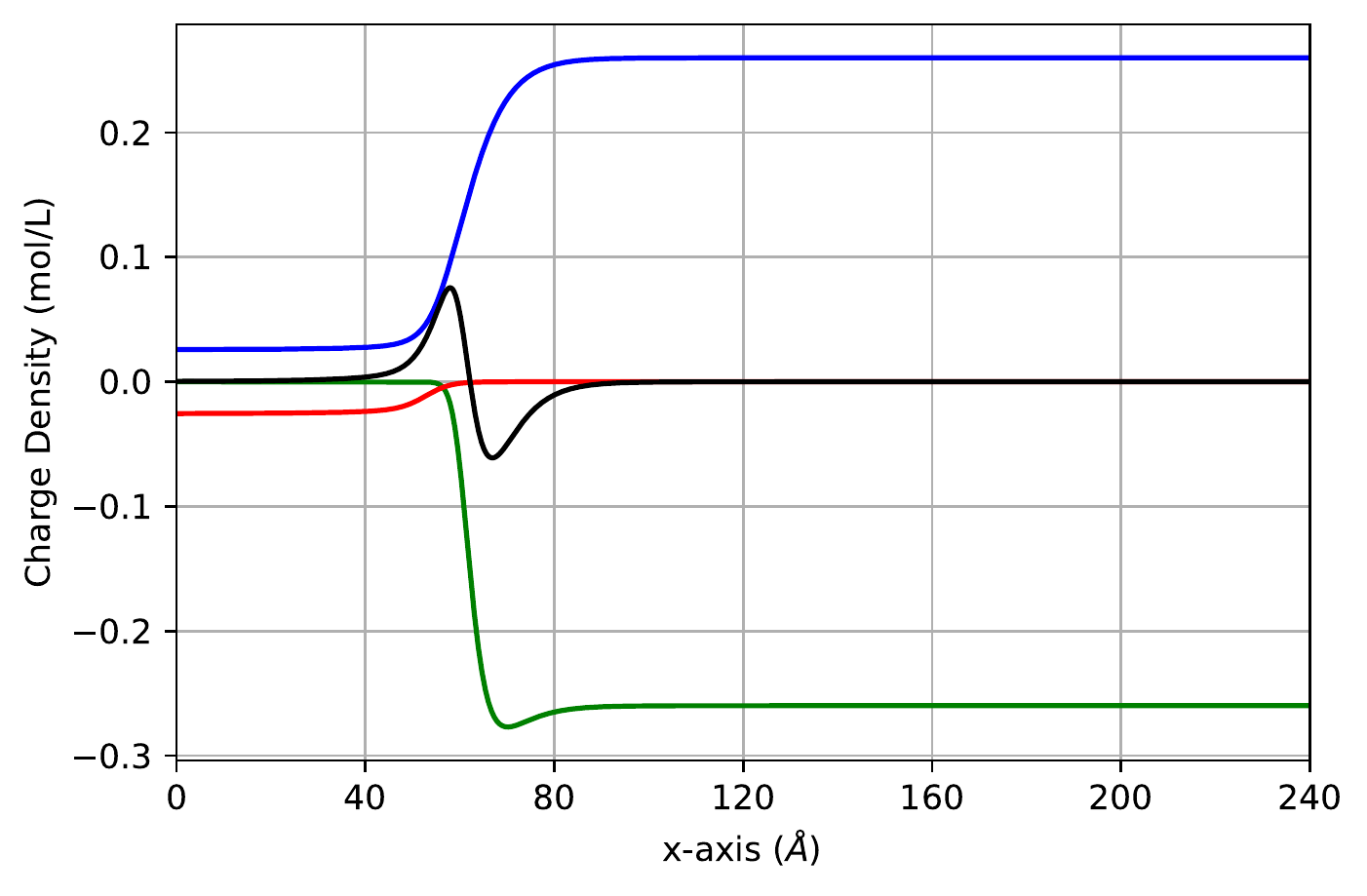}\\
    (c)\\
    \includegraphics[scale=0.55]{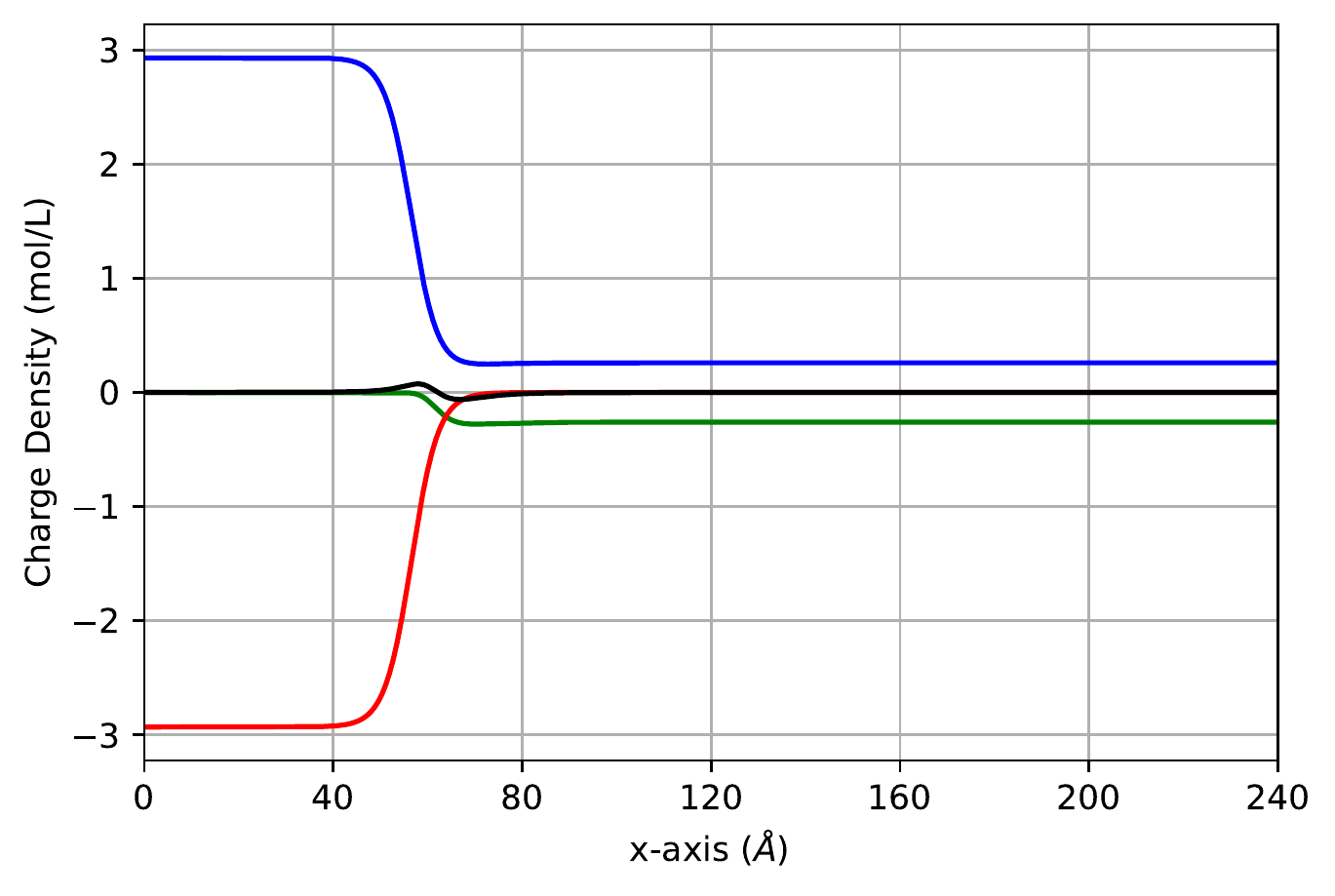}
    \caption{Heparin KCl Results. (a) Dimensionless Potential, (b) Unbound
    Charge Density (M), and (c) Total Charge Density (M) using parameters from
    Table \ref{sim_input_params}. In (b) and (c), the blue curve represents K,
    the green curve represents Cl, the red curve represents Heparin, and the
    black curve represents the net charge density.}
    \label{he_KCl_fig}
\end{figure}

The value of $\alpha$ to control the transition length of the permittivity and
total GAG concentration functions was qualitatively varied until the results in
Table~\ref{sim_out_eng} and the total charge density curves in Figure~\ref{he_KCl_fig}(c)
were in agreement with the corresponding data in \cite{sterling2021ion}.

\begin{table*}
  \caption{Parameters from \cite{sterling2021ion}.}
  \label{sim_params}
  \begin{ruledtabular}
  \begin{tabular}{llllllll}
    \textbf{GAG} & \textbf{Salt} & \textbf{Bulk Salt} & \textbf{GAG}
    & \textbf{Polymer} & $\varepsilon_{b}$ & $\varepsilon_{o}$ & T (K)\\
     & & \textbf{Conc.} & \textbf{Conc.} & \textbf{Length} & & & \\
     & & \textbf{(M)} & \textbf{(M)} & \textbf{(\textup{\AA})} & & & \\
    \hline
    Hyaluronan & NaCl & 0.28 & 0.51 & 129.7 & 50.7 & 60.4 & 310.15 \\
               & KCl  & 0.28 & 0.49 & 135.8 & 51.2 & 61.8 & 310.15 \\
    \hline
    Heparin    & NaCl & 0.27 & 2.78 & 119.3 & 37.5 & 59.1 & 310.15 \\
               & KCl  & 0.26 & 2.93 & 113.2 & 37.9 & 60.5 & 310.15 \\
  \end{tabular}
  \end{ruledtabular}
\end{table*}

\begin{table*}
  \caption{Input Parameters to Smoothly Varying Permittivity and
  Total GAG Concentration Model.}
  \label{sim_input_params}
  \begin{ruledtabular}
  \begin{tabular}{llllll}
    \textbf{Parameter} & \textbf{Notes} & \textbf{Hyaluronan}
    & \textbf{Hyaluronan} & \textbf{Heparin} & \textbf{Heparin} \\
     & & \textbf{NaCl} & \textbf{KCl}
     & \textbf{NaCl} & \textbf{KCl} \\
    \hline
    $\varepsilon_{r}$ & Dilute Water, $\varepsilon_{w}$ & $78.155$ & $78.155$ & $78.155$ & $78.155$ \\
    $\varepsilon_{1}(\hat{L})=\varepsilon_{S}$ & $\varepsilon_{o}/\varepsilon_{r}$ & $0.773$ & $0.791$
    & $0.756$ & $0.774$ \\
    $\varepsilon_{1}(0)=\varepsilon_{G}$ & $\varepsilon_{b}/\varepsilon_{r}$ & $0.649$ & $0.655$
    & $0.480$ & $0.485$ \\
    $\hat{\ell}$ & Polymer Length$/2/\lambda_{D}$ & $7.838$ & $8.206$ & $7.079$ & $6.592$ \\
    $\hat{L}$ & 240\textup{\AA}/$\lambda_{D}$ & $29.007$ & $29.007$ & $28.484$ & $27.951$ \\
    $r_{c}$ (\textup{\AA}) & \cite{sun2020analysis} & 1.62 & 1.95 & 1.62 & 1.95 \\
    $\hat{u}_{C}$ & Based on $r_{c}$ & 2.128 & 1.768 & 2.128 & 1.768 \\
    $r_{a}$ (\textup{\AA}) & Expected Value\cite{sun2020analysis} & 2.26 & 2.26 & 2.26 & 2.26 \\
     & Used Value & 0.226 & 0.226 & 0.226 & 0.226 \\
    $\hat{u}_{a}$ & Based on Expected $r_{a}$ & 1.525 & 1.525 & 1.525 & 1.525 \\
     & Based on Used $r_{a}$ & 15.251 & 15.251 & 15.251 & 15.251 \\
    $C_{0}$ & Salt Concentration & 0.28 & 0.28 & 0.27 & 0.26 \\
    $\widetilde{K}_{1}$ & Expected Value, Eqn (\ref{bind_to_K1}) & -0.115 & -0.192
    & $7.246\times 10^{-3}$ & $-6.310\times 10^{-3}$ \\
     & Used Value & 0.172 & 0.114 & $1.337\times 10^{-2}$ & $8.735\times 10^{-4}$ \\
    $\widetilde{c}$ & Salt Concentration/$C_{0}$ & 1 & 1 & 1 & 1 \\
    $\bar{c}$ & $\widetilde{c}\exp(\hat{u}_{c}/\varepsilon_{S})$ & $15.692$ & $9.350$
    & $16.671$ & $9.810$ \\
    $\bar{a}$ & $\bar{c}\exp[(\hat{u}_{a}-\hat{u}_{c})/\varepsilon_{S}]$ & $7.196$ & $6.881$
    & $7.515$ & $7.172$ \\
     & Used Value & $3.721\times 10^{8}$ & $2.380\times 10^{8}$ & $5.744\times 10^{8}$
     & $3.602\times 10^{8}$ \\
    $g_{0}$ & GAG Concentration/$C_{0}$ & $1.821$ & $1.75$ & $10.297$ & $11.269$ \\
    $\hat{\sigma}_{1}$ & -- & 0 & 0 & 0 & 0 \\
    $\hat{\sigma}_{2}$ & -- & 0 & 0 & 0 & 0 \\
    $\alpha$ & Expected Value & $\ll 1$ & $\ll 1$ & $\ll 1$ & $\ll 1$ \\
     & Used Value & 0.1 & 0.1 & 0.1 & 0.1 \\
  \end{tabular}
  \end{ruledtabular}
\end{table*}

As seen in Table~\ref{sim_out_eng}, our model predictions are well within the
standard deviations from the molecular simulation results. All of our predicted
energy values are within $0.014$ of the average values obtained from the molecular
simulations.

\begin{figure}
    \centering
    (a)\\
    \includegraphics[scale=0.55]{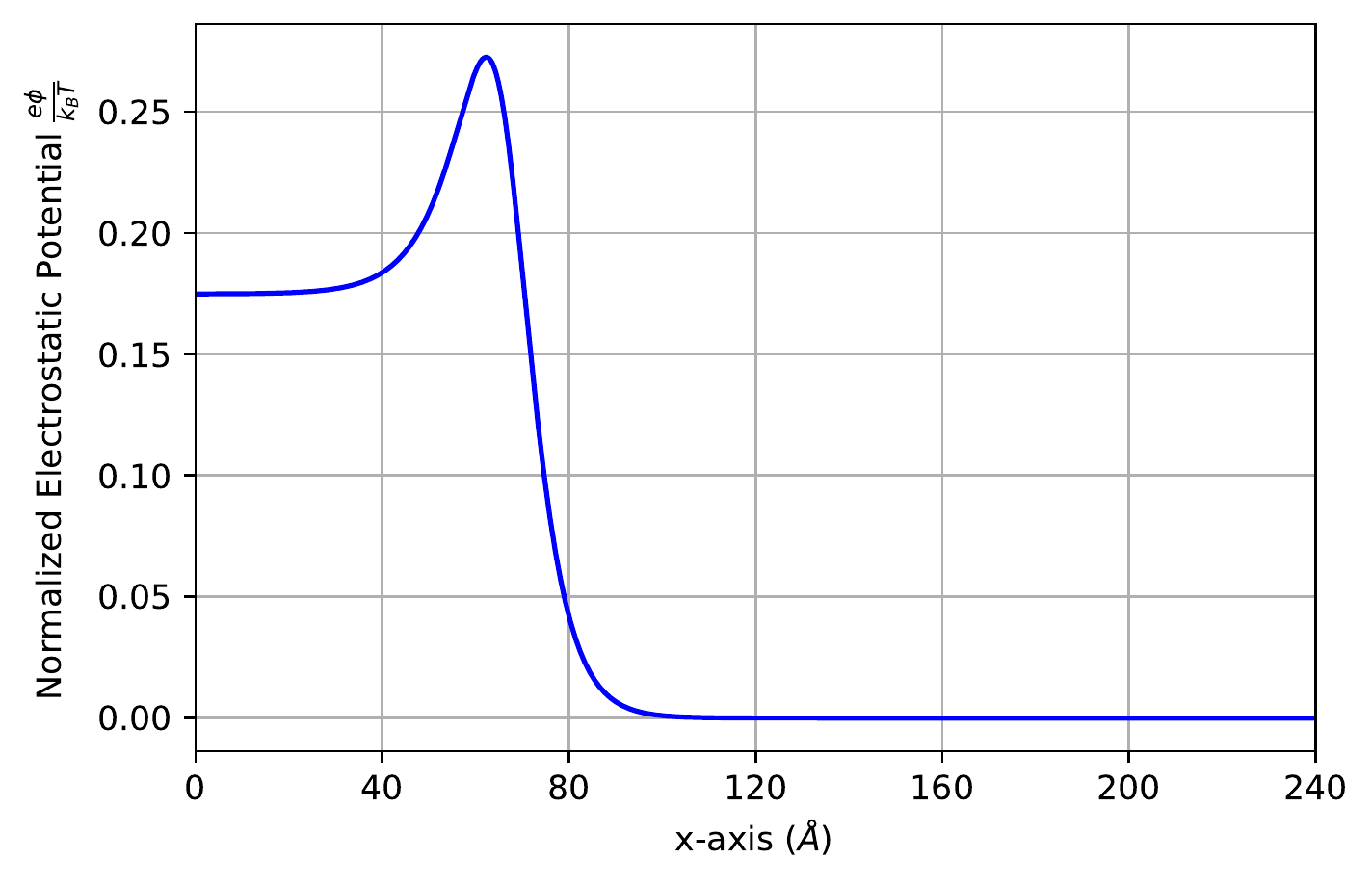}\\
    (b)\\
    \includegraphics[scale=0.55]{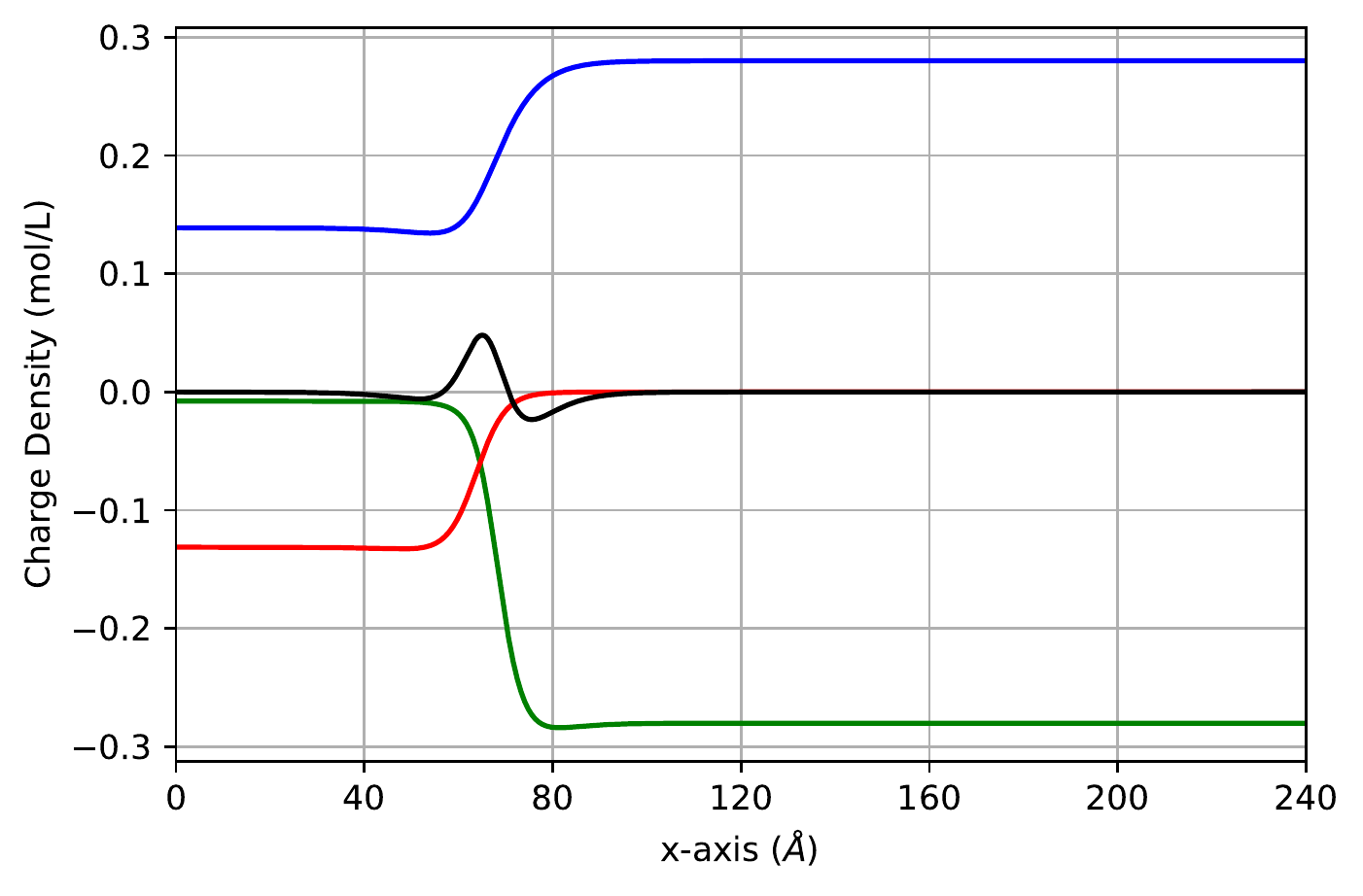}\\
    (c)\\
    \includegraphics[scale=0.55]{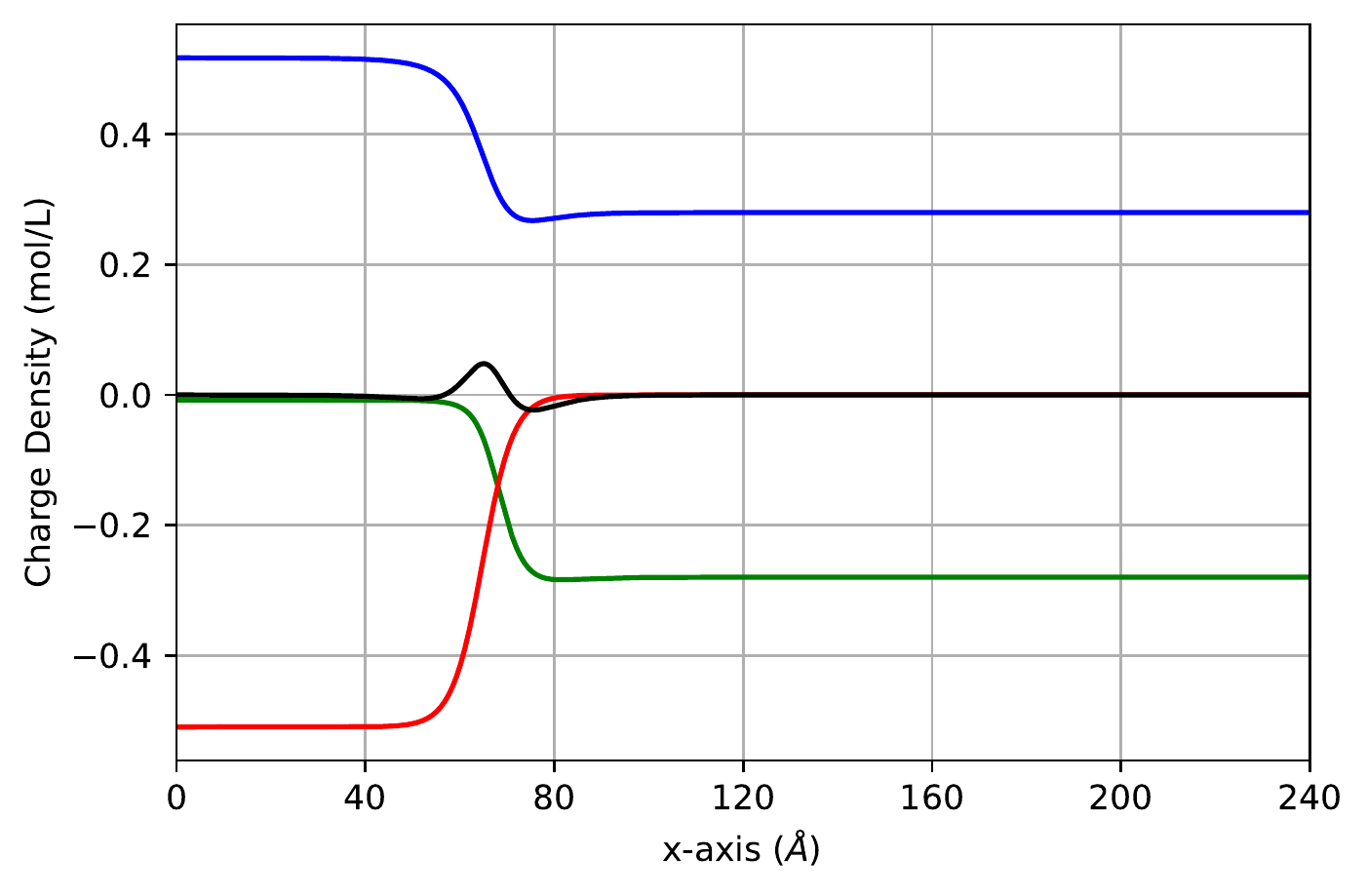}
    \caption{Hyaluronan NaCl Results. (a) Dimensionless Potential, (b) Unbound
    Charge Density (M), and (c) Total Charge Density (M) using parameters from
    Table \ref{sim_input_params}. In (b) and (c), the blue curve represents Na,
    the green curve represents Cl, the red curve represents Hyaluronan, and the
    black curve represents the net charge density.}
    \label{hy_NaCl_fig}
\end{figure}

\begin{figure}
    \centering
    (a)\\
    \includegraphics[scale=0.55]{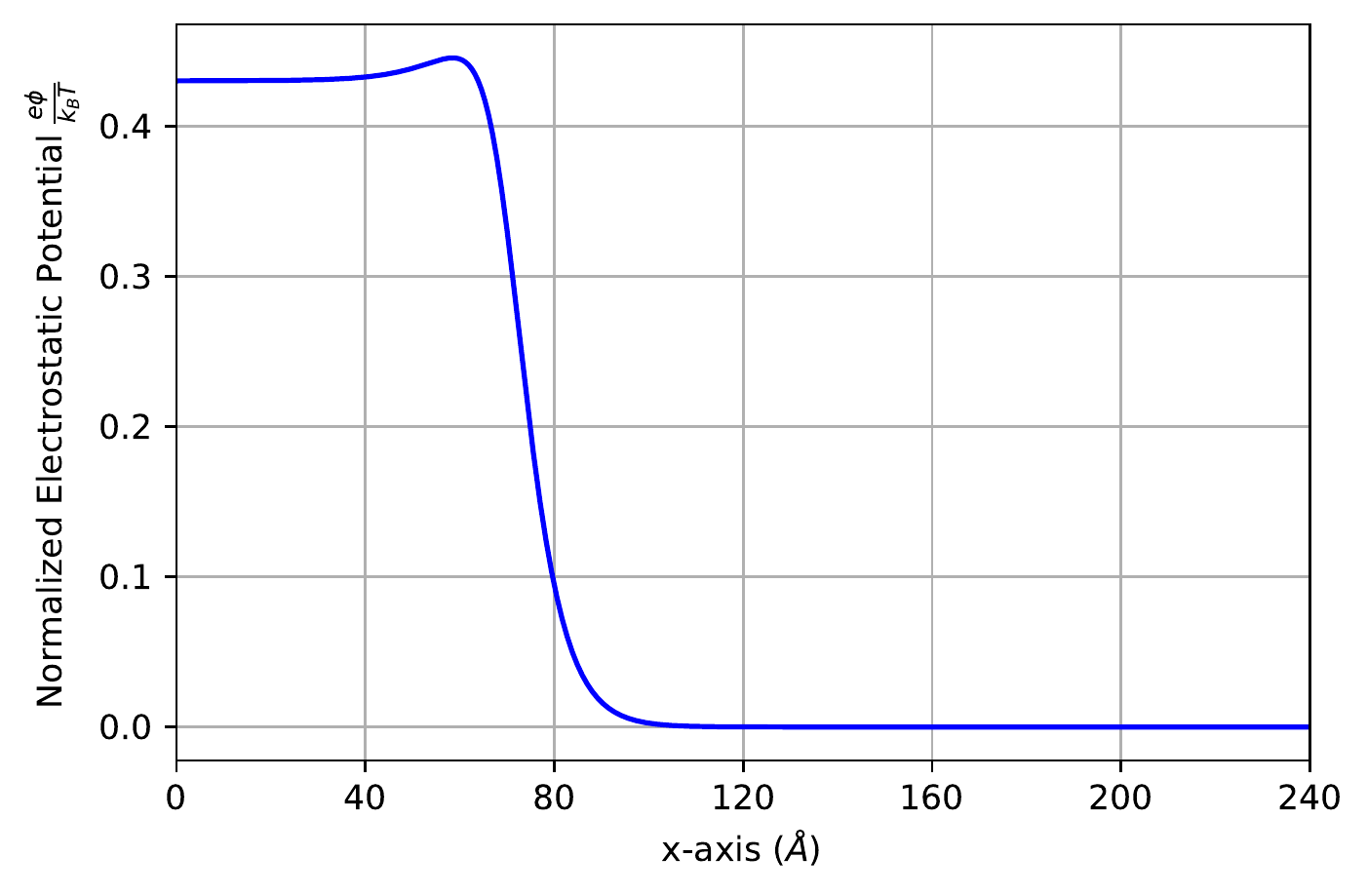}\\
    (b)\\
    \includegraphics[scale=0.55]{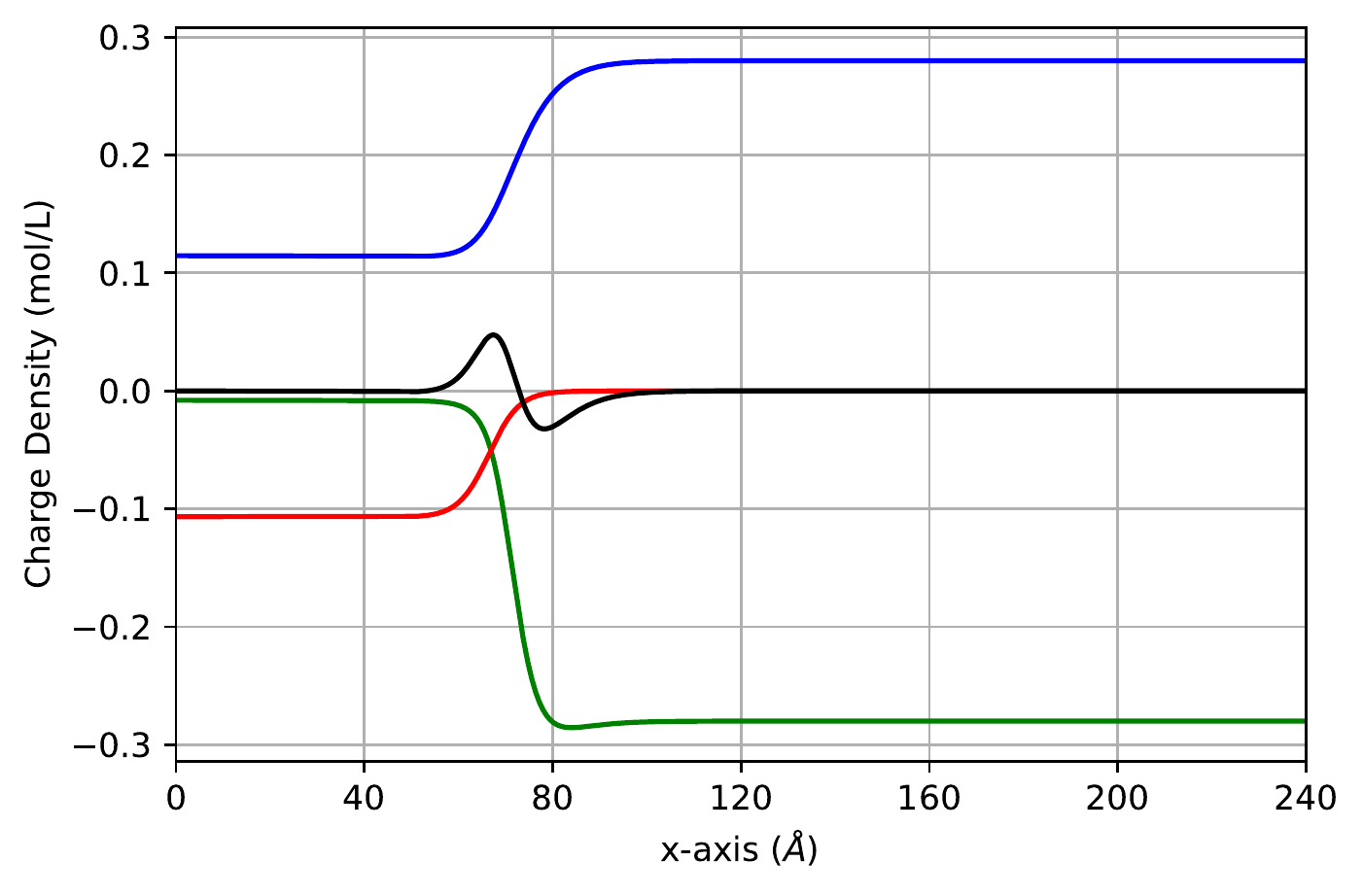}\\
    (c)\\
    \includegraphics[scale=0.55]{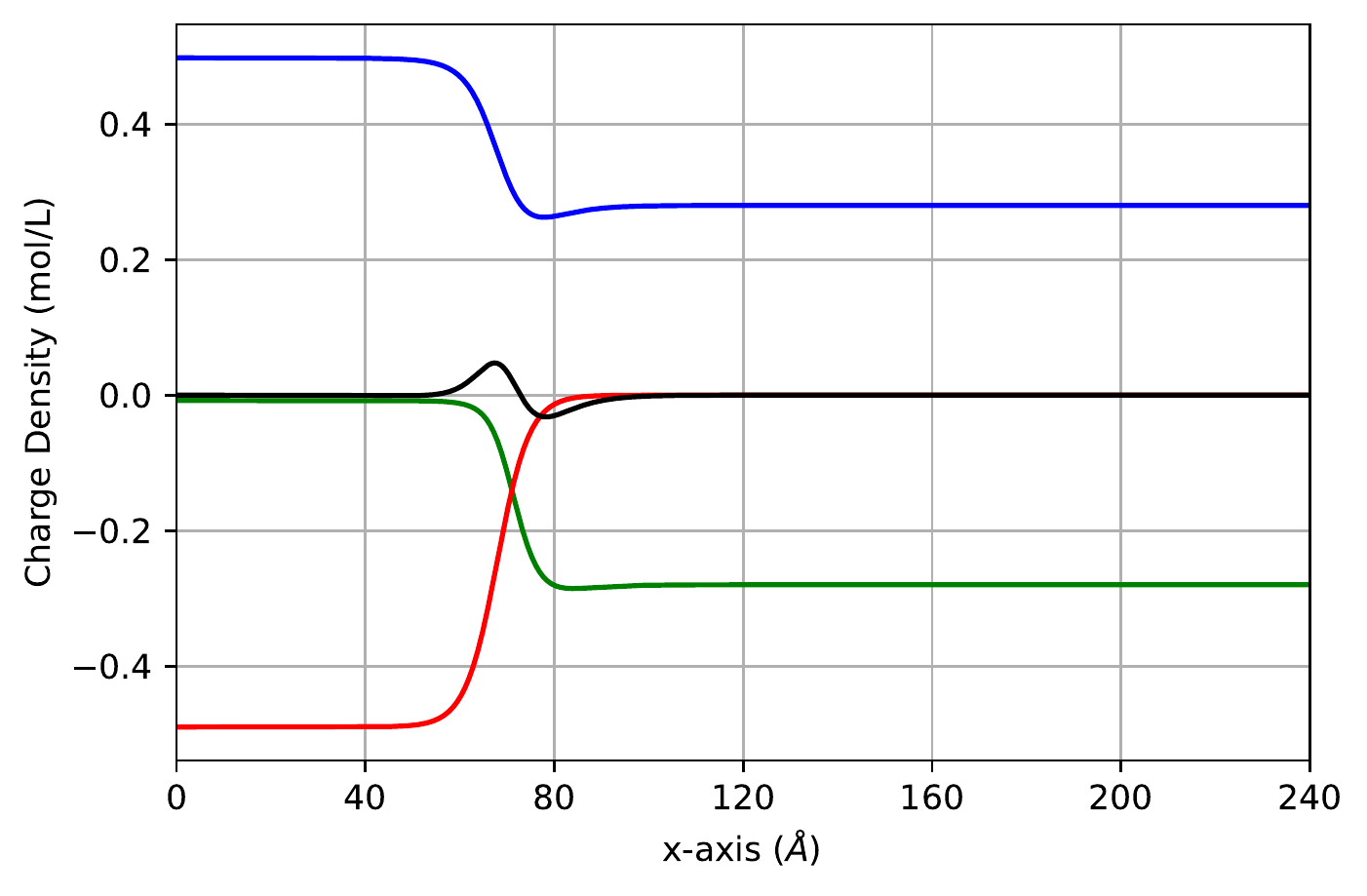}
    \caption{Hyaluronan KCl Results. (a) Dimensionless Potential, (b) Unbound
    Charge Density (M), and (c) Total Charge Density (M) using parameters from
    Table \ref{sim_input_params}. In (b) and (c), the blue curve represents K,
    the green curve represents Cl, the red curve represents Hyaluronan, and the
    black curve represents the net charge density.}
    \label{hy_KCl_fig}
\end{figure}

\begin{figure}
    \centering
    (a)\\
    \includegraphics[scale=0.55]{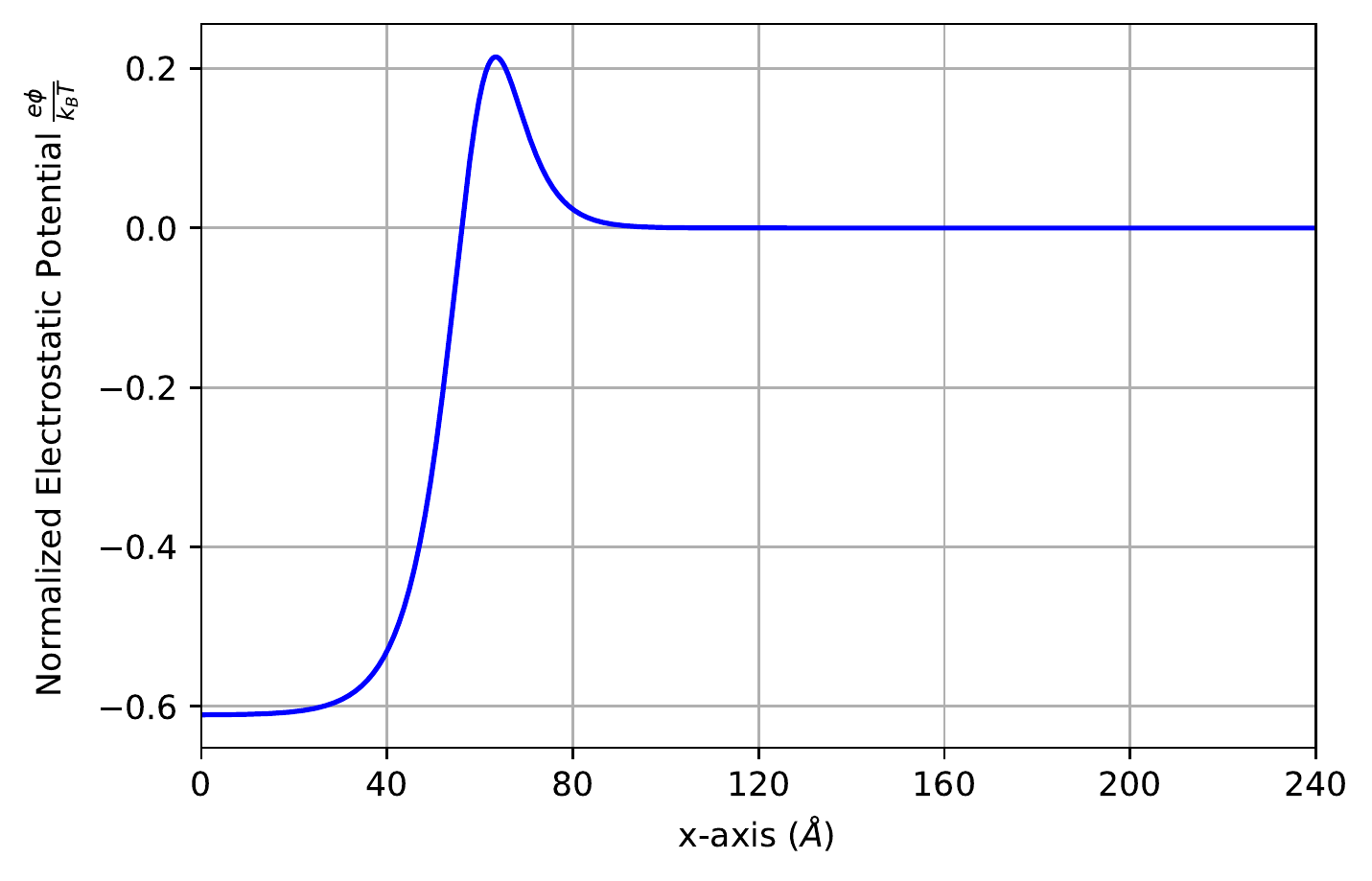}\\
    (b)\\
    \includegraphics[scale=0.55]{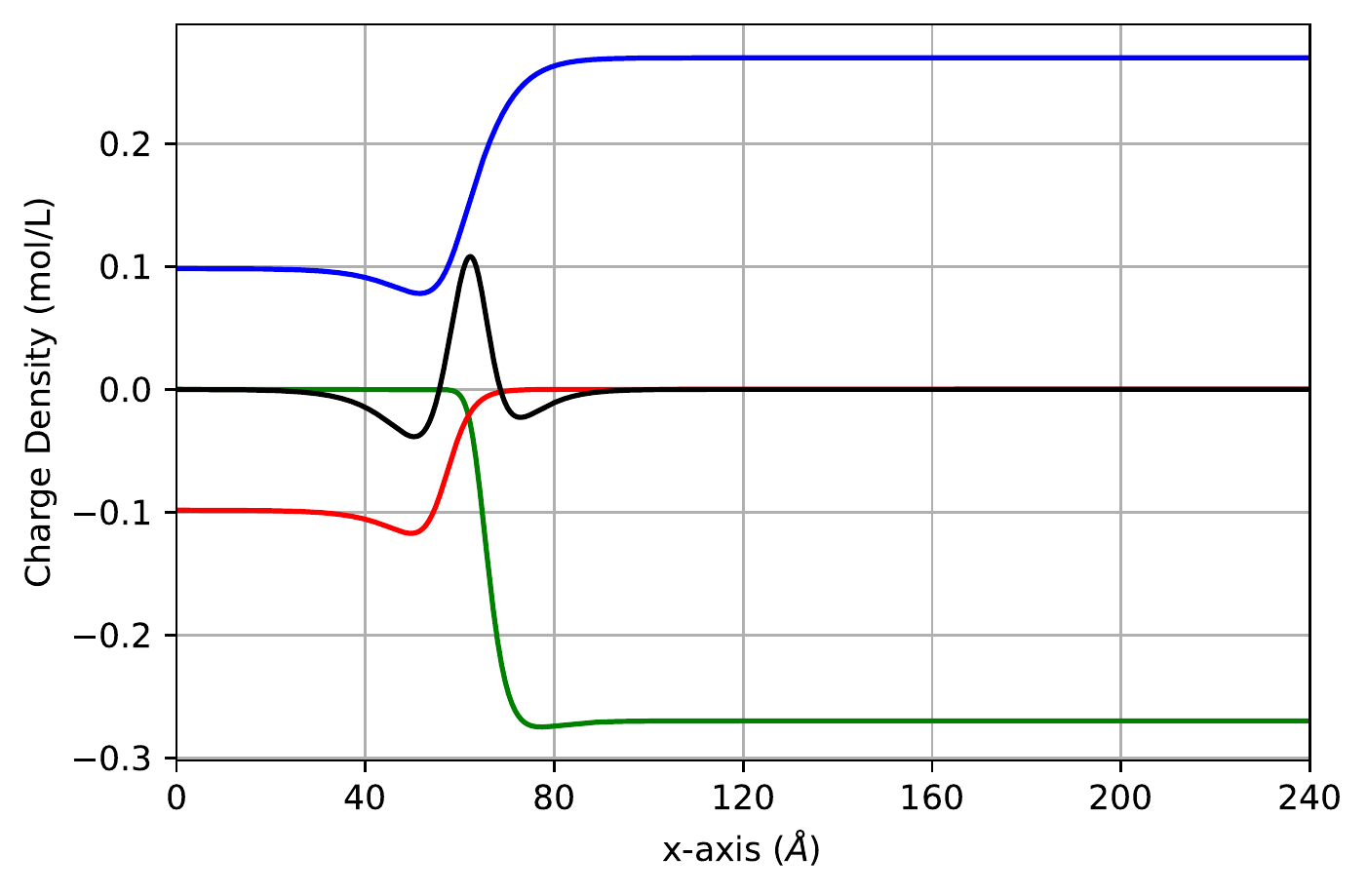}\\
    (c)\\
    \includegraphics[scale=0.55]{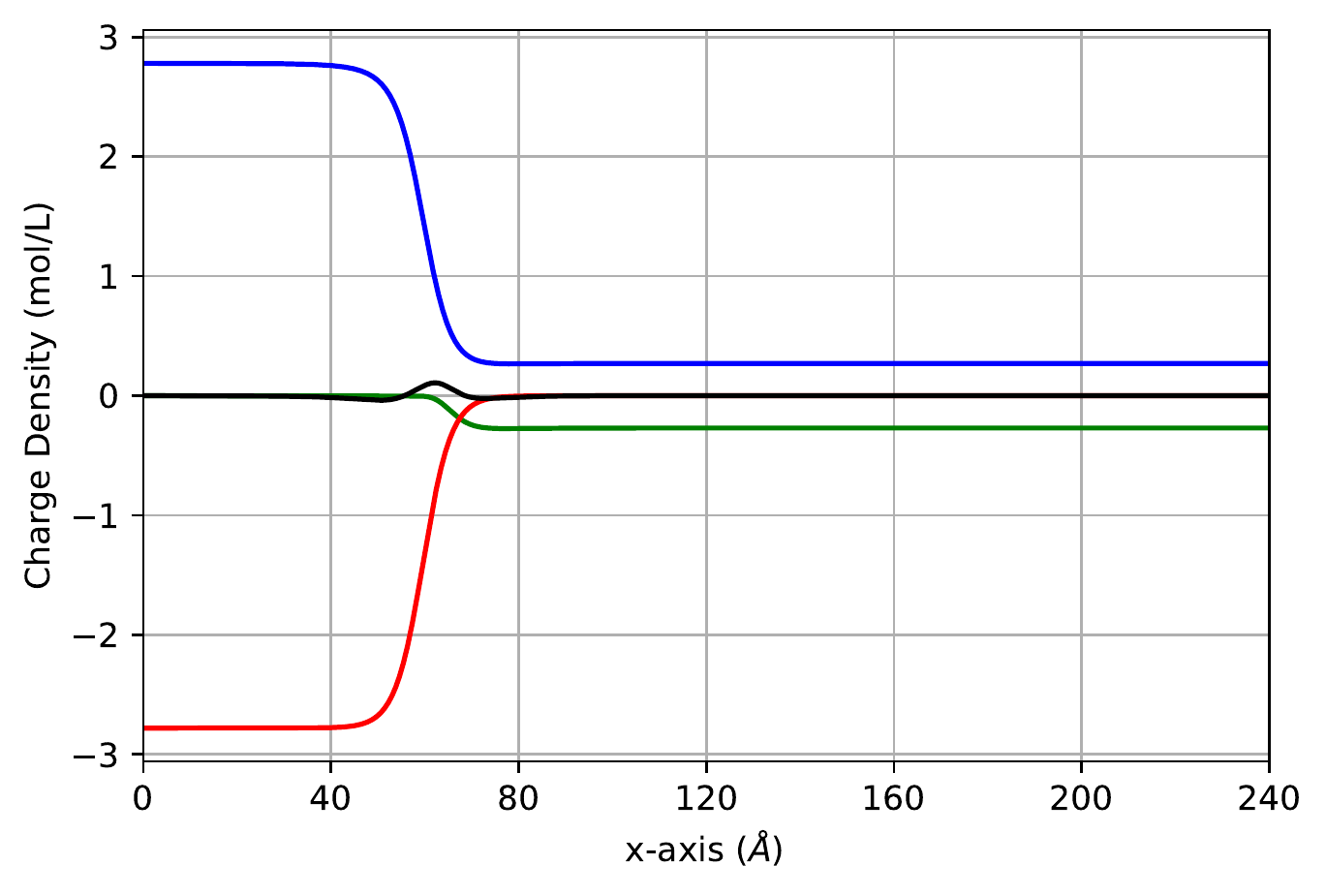}
    \caption{Heparin NaCl Results. (a) Dimensionless Potential, (b) Unbound
    Charge Density (M), and (c) Total Charge Density (M) using parameters from
    Table \ref{sim_input_params}. In (b) and (c), the blue curve represents Na,
    the green curve represents Cl, the red curve represents Heparin, and the
    black curve represents the net charge density.}
    \label{he_NaCl_fig}
\end{figure}

Plots of the predicted dimensionless potential, unbound charge density and total charge
density curves can be found in Figures \ref{hy_NaCl_fig}--\ref{he_KCl_fig}. For
Figures \ref{hy_KCl_fig} and \ref{he_KCl_fig} representing brushes with a potassium cation, the net charge density curves
exhibit a double-double layer of negative charge just outside of the brush edge and positive charge just inside of the brush edge. Observing the electrostatic potential, we see the same trend as expected in a dilute-limit where dielectric decrement effects are negligible: a negative unbound charge density at an $x$-location corresponds to a positive second derivative in the electrostatic potential, while a positive unbound charge density corresponds to a negative second derivative in the electrostatic potential. The molecular simulation data in \cite{sterling2021ion} exhibited the opposite trend, positive charge just outside of the brush and negative charge just inside of the brush. It is unknown at this time why this discrepancy exists, but this is the only substantial difference between our model predictions and the molecular simulation data. 

In contrast to the GAG brush results for the potassium cation, the net charge density curves in Figures \ref{hy_NaCl_fig} and \ref{he_NaCl_fig} in the presence of a sodium cation show more complex structure. The brushes exhibit a negative charge inside the edge of brush, then a positive charge in the transition region and a negative charge on the outside edge of the brush. This corresponds to an electrostatic potential 2nd-derivative that is positive-negative-positive meaning there is a substantial \textit{overshoot} of the potential rather than a smooth transition in the potential curve.

\begin{table*}
  \caption{Dimensionless Energy Results Molecular Simulation VS Model Prediction.}
  \label{sim_out_eng}
  \begin{ruledtabular}
  \begin{tabular}{llllll}
    \textbf{GAG} & \textbf{Salt} & & \textbf{Donnan} & \textbf{Born} & \textbf{Cation} \\
     & & & \textbf{Potential} & \textbf{Hydration} & \textbf{Binding} \\
     & & & & \textbf{Energy} & \textbf{Energy} \\
    \hline
    Hyaluronan  & NaCl & Molecular Simulation\cite{sterling2021ion} & 0.17$\pm0.06$
    & 0.53$\pm0.2$  & -1.32$\pm0.2$  \\
                &      & Model Prediction                             & 0.175 & 0.527 & -1.316 \\
    \hline
    Hyaluronan  & KCl  & Molecular Simulation\cite{sterling2021ion} & 0.44$\pm0.03$
    & 0.46$\pm0.2$  & -1.46$\pm0.2$  \\
                &      & Model Prediction                             & 0.430 & 0.463 & -1.469 \\
    \hline
    Heparin     & NaCl & Molecular Simulation\cite{sterling2021ion} & -0.62$\pm0.27$
    & 1.62$\pm0.6$  & -3.35$\pm0.6$  \\
                &      & Model Prediction                             & -0.611 & 1.621 & -3.342 \\
    \hline
    Heparin     & KCl  & Molecular Simulation\cite{sterling2021ion} & 0.96$\pm0.14$
    & 1.36$\pm0.5$  & -4.73$\pm0.5$  \\
                &      & Model Prediction                             & 0.946 & 1.362 & -4.730 \\
  \end{tabular}
  \end{ruledtabular}
\end{table*}

\section{Extending the Model}
The model as derived above focused on a monovalent salt solution consisting of
one cation and one anion. This can be expanded to incorporate a second cation by
modifying equation (\ref{mono-salt}) to be
\beqa\label{second-cat}
-\frac{d}{dx}\left(\varepsilon\frac{d\phi}{dx}\right) 
&= N_{A}e\left(\left[C_{1}^{+}\right] + \left[C_{2}^{+}\right] - \left[A^{-}\right]
- \left[G^{-}\right]\right),\nonumber\\ &\qquad \text{in} \ \Omega,
\eeqa
where $\left[C_{1}^{+}\right]$ represents the cation 1 concentration and
$\left[C_{2}^{+}\right]$ represents the cation 2 concentration. There are now two
ion pairs, which leads to the modified equation (\ref{ion-pair}) as
\beq
\left[GC_{1}\right] = \frac{\left[C_{1}^{+}\right]}{K_{1}}\left[G^{-}\right],\qquad
\left[GC_{2}\right] = \frac{\left[C_{2}^{+}\right]}{K_{2}}\left[G^{-}\right],
\eeq
where $K_{1}$ and $K_{2}$ are the dissociation constants for cations 1 and
2, respectively.

The total GAG concentration, $\left[G^{-}\right]_{0}$, is now the sum of the unbound
GAG concentration, $\left[G^{-}\right]$, and the concentrations of the two bound
ion pairs, $\left[GC_{1}\right]$ and $\left[GC_{2}\right]$, i.e., $\left[G^{-}\right]_{0} = \left[G^{-}\right] + \left[GC_{1}\right] + \left[GC_{2}\right]$. From this, we infer that:
\beq
\left[G^{-}\right] = \frac{\left[G^{-}\right]_{0}}{1
+ {\left[C_{1}^{+}\right]}/{K_{1}} + {\left[C_{2}^{+}\right]}/{K_{2}}} \ .
\eeq

\subsection{Piecewise Constant Model}
Following the same procedure and defintions from section \ref{pw_const}, we arrive
at the modified ODE in the Salt region
\beq\label{salt_final2}
-\frac{d^{2}y}{d\hat{x}^{2}}= \frac{1}{\varepsilon_{S}}\left[\widetilde{c}_{1}e^{-y}
+\widetilde{c}_{2}e^{-y}-\left(\widetilde{c}_{1}+\widetilde{c}_{2}\right)e^{y}\right]
\qquad \text{in} \ \Omega_{S}.
\eeq
Similarly, the modified ODE in the GAG region is
\begin{widetext}
\beqa\label{gag_final2}
-\frac{d^{2}y}{d\hat{x}^{2}}&=& \frac{1}{\varepsilon_{G}}
\bigg\{\widetilde{c}_{1}\exp\left(-y-\hat{u}_{c1}
\left[\frac{1}{\varepsilon_{G}}-\frac{1}{\varepsilon_{S}}\right]\right)
+\widetilde{c}_{2}\exp\left(-y-\hat{u}_{c2}
\left[\frac{1}{\varepsilon_{G}}-\frac{1}{\varepsilon_{S}}\right]\right)\nonumber\\
&&-\left(\widetilde{c}_{1}+\widetilde{c}_{2}\right)\exp\left(y-\hat{u}_{a}
\left[\frac{1}{\varepsilon_{G}}-\frac{1}{\varepsilon_{S}}\right]\right)\bigg\}\nonumber\\
&&-\frac{1}{\varepsilon_{G}}\left\{\frac{\bar{g}}{1+\frac{\widetilde{c}_{1}}{\widetilde{K}_{1}}
\exp\left(-y-\hat{u}_{c1}
\left[\frac{1}{\varepsilon_{G}}-\frac{1}{\varepsilon_{S}}\right]\right)
+\frac{\widetilde{c}_{2}}{\widetilde{K}_{2}}
\exp\left(-y-\hat{u}_{c2}
\left[\frac{1}{\varepsilon_{G}}-\frac{1}{\varepsilon_{S}}\right]\right)}\right\} \qquad
\text{in} \ \Omega_{G}.
\eeqa
\end{widetext}
The boundary conditions remain unchanged from equations (\ref{bc1-piecewise})--(\ref{bc4-piecewise}).

\subsection{Smooth Model}
Following the same procedure and definitions from section \ref{smooth}, we arrive
at the modified ODE
\begin{widetext}
\beqa\label{smooth_final2}
-\varepsilon_{1}(\hat{x})\frac{d^{2}y}{d\hat{x}^{2}}
-\frac{d \varepsilon_{1}(\hat{x})}{d\hat{x}}\frac{dy}{d\hat{x}}
&=& \bar{c}_{1}\exp\left(-y - \frac{\hat{u}_{c1}}{\varepsilon_{1}(\hat{x})}\right)
+ \bar{c}_{2}\exp\left(-y - \frac{\hat{u}_{c2}}{\varepsilon_{1}(\hat{x})}\right)
- \bar{a}\exp\left(y - \frac{\hat{u}_{a}}{\varepsilon_{1}(\hat{x})}\right)\nonumber\\
&&- \dfrac{\bar{g}(\hat{x})}{1+(\bar{c}_{1}/\widetilde{K}_{1})
\exp\left(-y - \frac{\hat{u}_{c1}}{\varepsilon_{1}(\hat{x})}\right)
+(\bar{c}_{2}/\widetilde{K}_{2})=
\exp\left(-y - \frac{\hat{u}_{c2}}{\varepsilon_{1}(\hat{x})}\right)} \qquad \text{in} \ \Omega,
\eeqa
\end{widetext}
where
\beq
\bar{a} = \bar{c}_{1}\exp\left(\frac{\hat{u}_{a}-\hat{u}_{c1}}{\varepsilon_{S}}\right)
+\bar{c}_{2}\exp\left(\frac{\hat{u}_{a}-\hat{u}_{c2}}{\varepsilon_{S}}\right).
\eeq
The boundary conditions remain unchanged from equations (\ref{bc1-smooth})--(\ref{bc2-smooth}).

\subsection{A Time-Dependent Model}
The ODE model focuses on the steady state solution of the GAG/salt system. A complete mathematical
description of the system requires consideration of the transient solution prior to
reaching steady state. We lay out a foundation for a time-dependent model that we can build on and
continue to develop. We begin by assuming a constant permittivity allowing us to neglect the
Born energy term for the time being.

For a species, $\left[C_{i}\right]$, undergoing diffusion, chemical reaction, and drift due to an
external force, the conservation equation is
\beq\label{cons_eq}
\frac{\partial \left[C_{i}\right]}{\partial t} + \frac{\partial J_{i}}{\partial x} = r_{i},
\eeq
where $r_{i}$ is the net rate of production of the ion species per unit volume. Flux $J_{i}$ depends on the drift velocity, $v_{d,i}$ and the diffusion coefficient, $D_{i}$, by
\beq
J_{i} = v_{d,i}\left[C_{i}\right] - D_{i}\frac{\partial \left[C_{i}\right]}{\partial x}.
\eeq
The drift velocity is related to the external force on the species via the mobility,
which by the Stokes-Einstein relation is
\beq
v_{d,i} = \frac{D_{i}}{k_{B}T}F_{i}.
\eeq
Here, the external force, $F_{i}$, represents the Coulombic force that drives ion electrophoresis and is proportional to the electric field. Using equation (\ref{def_pot}) this force can be written
\beq\label{force_pot}
F_{i} = -z_{i}e\frac{\partial\phi}{\partial x}.
\eeq
Combining equations (\ref{cons_eq})--(\ref{force_pot}) yields
\beq
\frac{\partial \left[C_{i}\right]}{\partial t}
+ \frac{\partial}{\partial x}\left(-D_{i}\frac{\partial\left[C_{i}\right]}{\partial x}
- \frac{z_{i}e}{k_{B}T}D_{i}\frac{\partial\phi}{\partial x}\left[C_{i}\right]\right) = r_{i}.
\eeq
To obtain a dimensionless form, we use the same scaling parameters defined previously for
concentration, length, and potential. Scale all diffusion constants with some value, $D_{0}$, and define the dimensionless diffusion constants as
\beq
d_{i} = \frac{D_{i}}{D_{0}}.
\eeq
We can then define a time scale using $\lambda_{D}$ and $D_{0}$ to obtain
\beq
\hat{t} = \frac{t}{(\lambda_{D}^{2}/D_{0})}.
\eeq
Finally, the net rate of production is non-dimensionalized as
\beq
\hat{r}_{i} = \frac{\lambda_{D}^{2}}{D_{0}C_{0}}r_{i}.
\eeq
This yields the dimensionless PDE
\beq
\frac{\partial c_{i}}{\partial\hat{t}}
+ \frac{\partial}{\partial \hat{x}}\left(-d_{i}\frac{\partial c_{i}}{\partial\hat{x}}
- z_{i}d_{i}\frac{\partial y}{\partial\hat{x}}c_{i}\right) = \hat{r}_{i}.
\eeq

Considering the negatively charged GAG brush in a salt solution consisting of two cations and
one anion, the equation for the potential is the same as equation (\ref{second-cat}) with the
ordinary derivatives replaced with partial derivatives and non-dimensionalized as before to obtain
\beq\label{pde_pot}
-\frac{\partial^2 y}{\partial\hat{x}^2}
= c_{1} + c_{2} - a - g.
\eeq
In this system, the net rates of production are governed by the binding chemical reactions
\[-k_{1}\left[G^{-}\right]\left[C_{1}^{+}\right]+k_{-1}\left[GC_{1}\right]\]
and
\[-k_{2}\left[G^{-}\right]\left[C_{2}^{+}\right]+k_{-2}\left[GC_{2}\right].\]
The forward reaction rate constants, $k_{1}$ and $k_{2}$, can be non-dimensionalized as
\bseq
\beq
\hat{k}_{1} = \frac{\lambda_{D}^{2}C_{0}}{D_{0}}k_{1},
\eeq
\beq
\hat{k}_{2} = \frac{\lambda_{D}^{2}C_{0}}{D_{0}}k_{2}.
\eeq
\eseq
The backward reaction rate constants, $k_{-1}$ and $k_{-2}$, can be non-dimensionalized as
\bseq
\beq
\hat{k}_{-1} = \frac{\lambda_{D}^{2}}{D_{0}}k_{-1},
\eeq
\beq
\hat{k}_{-2} = \frac{\lambda_{D}^{2}}{D_{0}}k_{-2}.
\eeq
\eseq
Combined with the dimensionless Poisson equation for the electrostatic potential (\ref{pde_pot}), the dimensionless 
time-dependent species conservation equations are known as the Poisson-Nernst-Planck (PNP) equations, 
\bseq
\beq
\frac{\partial c_{1}}{\partial\hat{t}}
+\frac{\partial}{\partial\hat{x}}\left(-d_{c1}\frac{\partial c_{1}}{\partial\hat{x}}
-d_{c1}\frac{\partial{y}}{\partial\hat{x}}c_{1}\right)
=-\hat{k}_{1}c_{1}g+\hat{k}_{-1}\left[gc_{1}\right],
\eeq
\beq
\frac{\partial c_{2}}{\partial\hat{t}}
+\frac{\partial}{\partial\hat{x}}\left(-d_{c2}\frac{\partial c_{2}}{\partial\hat{x}}
-d_{c2}\frac{\partial{y}}{\partial\hat{x}}c_{2}\right)
=-\hat{k}_{2}c_{2}g+\hat{k}_{-2}\left[gc_{2}\right],
\eeq
\beq
\frac{\partial a}{\partial\hat{t}}
+\frac{\partial}{\partial\hat{x}}\left(-d_{a}\frac{\partial a}{\partial\hat{x}}
+d_{a}\frac{\partial{y}}{\partial\hat{x}}a\right)
=0,
\eeq
\beq
\frac{\partial g}{\partial\hat{t}} = -\hat{k}_{1}c_{1}g+\hat{k}_{-1}\left[gc_{1}\right]
-\hat{k}_{2}c_{2}g+\hat{k}_{-2}\left[gc_{2}\right],
\eeq
\beq
\frac{\partial \left[gc_{1}\right]}{\partial\hat{t}}
= \hat{k}_{1}c_{1}g-\hat{k}_{-1}\left[gc_{1}\right],
\eeq
\beq
\frac{\partial \left[gc_{2}\right]}{\partial\hat{t}}
= \hat{k}_{2}c_{2}g-\hat{k}_{-2}\left[gc_{2}\right].
\eeq
\eseq

If there are no surface charge densities on either boundary and no sources or sinks for the
cations or anions, then the boundary conditions for all of the dependent variables are homogeneous
Neumann boundary conditions.

\subsection{Transient Results}
At this point, we want to examine a few example scenarios, look at any interesting transient
responses and compare the steady state solution of the time-dependent model to the ODE model.
To that end, we defined three different sets of initial conditions shown in Table \ref{pde_input_params}
along with Neumann boundary conditions and utilized COMSOL Multiphysics 6.0 to calculate the numerical
solutions to our time-dependent model. To keep consistent with our ODE model defintiions, we define
the GAG brush region to be $0\leq\hat{x}\leq 10$ and the salt region to be $10\leq\hat{x}\leq 20$.

\begin{table}
  \caption{Input Parameters and Initial Conditions for the Time-Dependent Model.}
  \label{pde_input_params}
  \begin{ruledtabular}
  \begin{tabular}{llll}
    \textbf{Parameter/}
    & \textbf{Scenario 1} & \textbf{Scenario 2} 
    & \textbf{Scenario 3} \\
    \textbf{Initial Condition} & & & \\
    \hline
    $d_{c1}$ & 1 & 1 & 1 \\
    $d_{c2}$ & 1 & 1 & 1 \\
    $d_{a}$ & 1 & 1 & 1 \\
    $\hat{k}_{1}$ & 0.5 & 0.5 & 0.5 \\
    $\hat{k}_{-1}$ & 5 & 5 & 5 \\
    $\hat{k}_{2}$ & 0.5 & 0.5 & 0.5 \\
    $\hat{k}_{-2}$ & 0.5 & 0.5 & 0.5 \\
    $y(\hat{x}, 0)$ & 0 & see Fig \ref{pde_init_cond} & see Fig \ref{pde_init_cond} \\
    $c_{1}(\hat{x}, 0)$ & $5*(\hat{x}<10)$ & see Fig \ref{pde_init_cond} & see Fig \ref{pde_init_cond} \\
    $c_{2}(\hat{x}, 0)$ & $(\hat{x}>10)$ & $10*(\hat{x}<1)$ & $10*(\hat{x}>19)$ \\
    $a(\hat{x}, 0)$ & $(\hat{x}>10)$ & $10*(\hat{x}<1)$ & $10*(\hat{x}>19)$ \\
    $g(\hat{x}, 0)$ & $5*(\hat{x}<10)$ & see Fig \ref{pde_init_cond} & see Fig \ref{pde_init_cond} \\
    $[gc_{1}](\hat{x}, 0)$ & 0 & see Fig \ref{pde_init_cond} & see Fig \ref{pde_init_cond} \\
    $[gc_{2}](\hat{x}, 0)$ & 0 & 0 & 0 \\
  \end{tabular}
  \end{ruledtabular}
\end{table}
\begin{figure*}
    \includegraphics[scale=0.53]{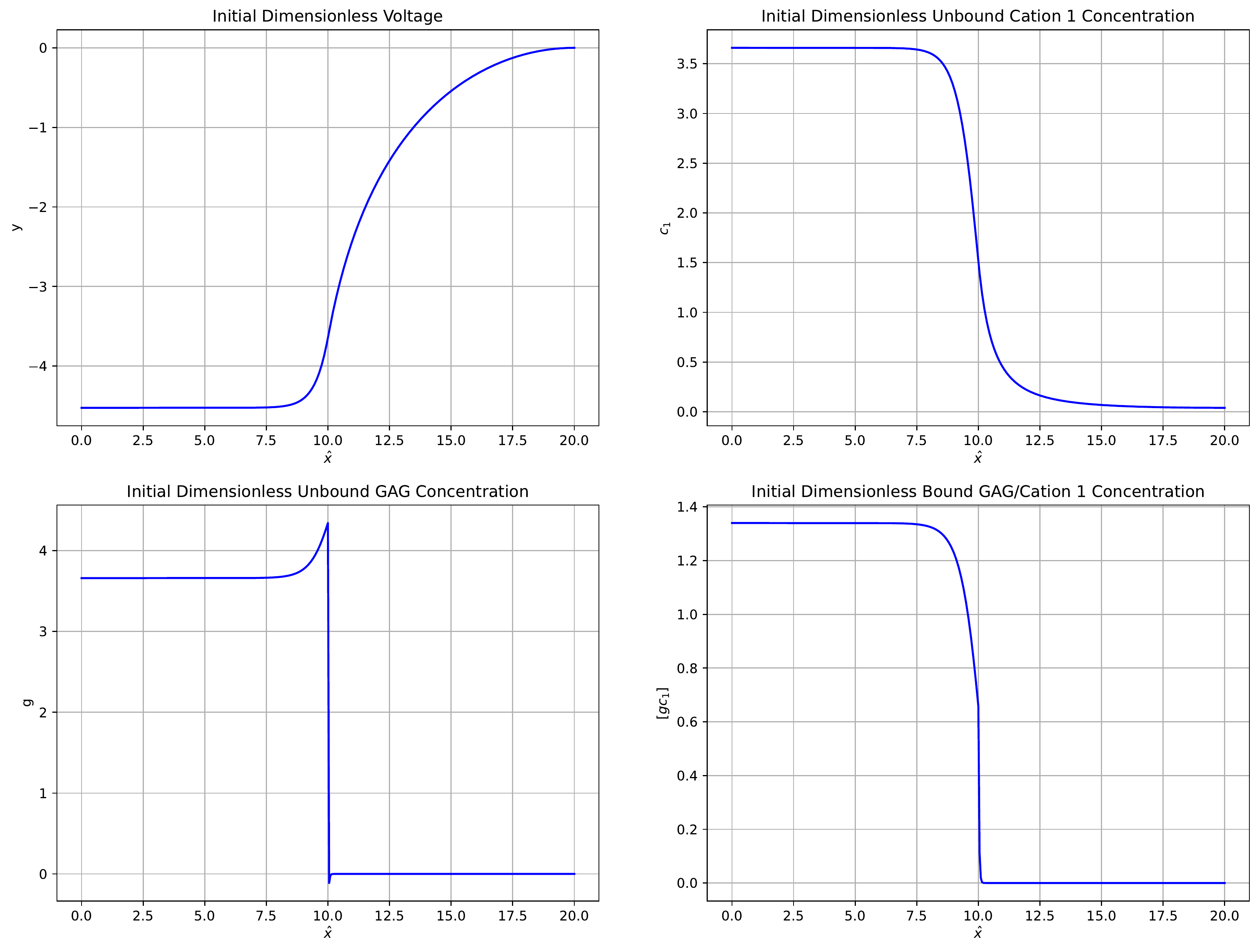}
    \caption{Initials conditions for voltage (top left), unbound cation 1 concentration (top right),
    unbound GAG concentration (bottom left), and bound GAG/cation 1 concentration (bottom right).}
    \label{pde_init_cond}
\end{figure*}

\begin{figure}
    \centering
    \includegraphics[scale=0.62]{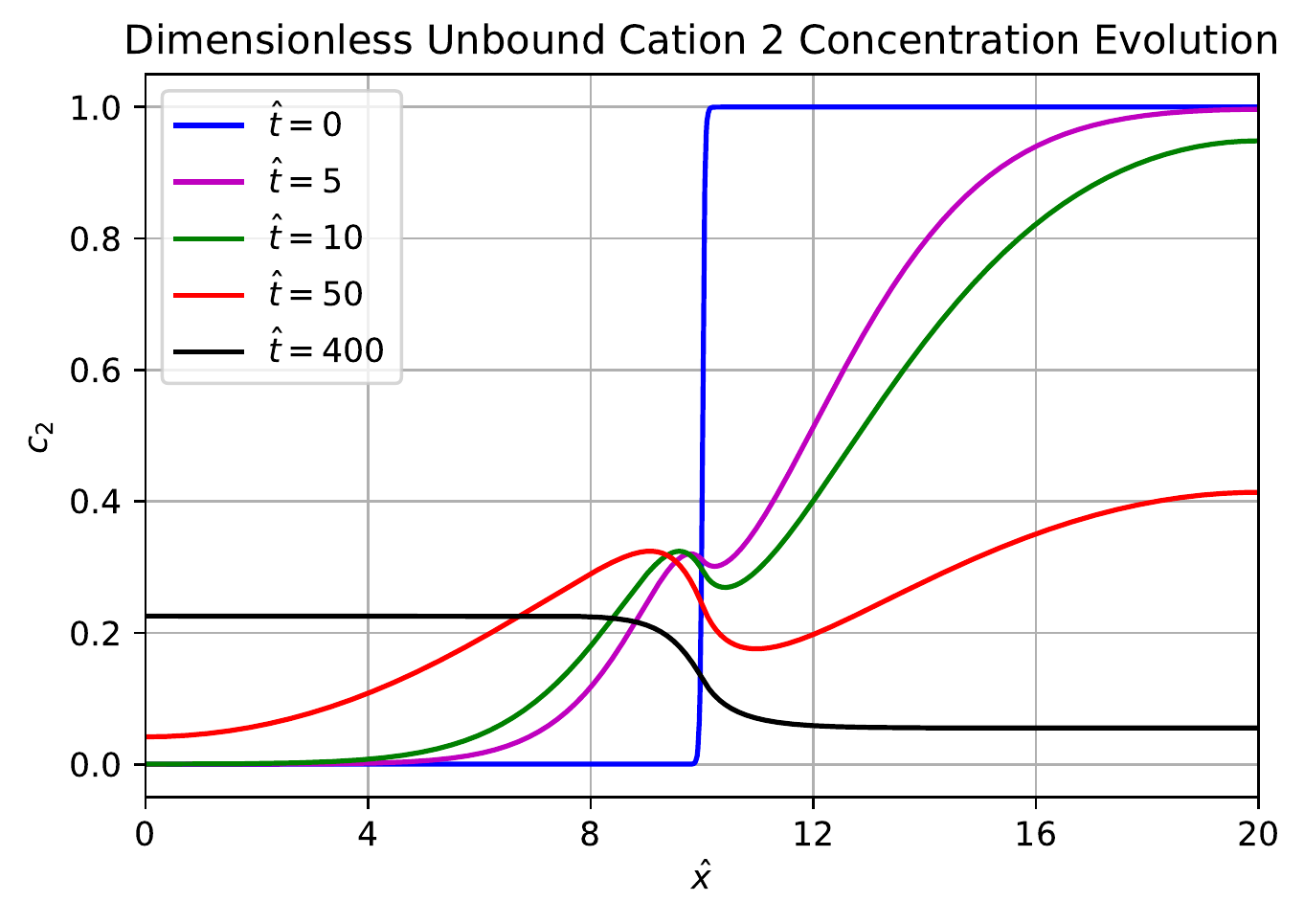}
    \caption{Time evolution for the dimensionless unbound cation 2 concentration.}
    \label{scen01_c2evol}
\end{figure}

We found that the dimensionless unbound cation 2 concentration for scenario 1 had the most
interesting transient response. From Table \ref{pde_input_params}, we chose an initial
condition for cation 2 such that there was no initial concentration in the GAG brush region
and a uniform concentration in the salt region. As seen in Figure \ref{scen01_c2evol},
unbound cation 2 ions want to leave the salt region and bind with the GAG ions in the brush
region. During the transient period, the curve changes from being convex in the brush region
and concave in the salt region to being concave in the brush region and convex in the salt region.

The initial condition for all 3 scenarios were chosen in such a way that the steady state
solutions should all converge to the same final equilibria. This was achieved by using the same total amount
of concentration for each ion when integrated over the whole domain. Table \ref{pde_result}
shows the boundary values of all the independent variables at $\hat{t}=400$, which was chosen
large enough to allow the system to reach steady state. There is good numerical agreement between
the 3 scenarios. This result gives us confidence that the steady state solution
is numerically stable.

\begin{table}
  \caption{Steady State Result Comparison at $\hat{t}=400$.}
  \label{pde_result}
  \begin{ruledtabular}
  \begin{tabular}{llll}
    \textbf{Parameter}
    & \textbf{Scenario 1} & \textbf{Scenario 2} 
    & \textbf{Scenario 3} \\
    \hline
    $y(0, 400)$ & -1.41 & -1.41 & -1.42 \\
    $y(20, 400)$ & 0 & 0 & 0 \\
    $c_{1}(0, 400)$ & 3.21 & 3.20 & 3.21 \\
    $c_{1}(20, 400)$ & 0.78 & 0.78 & 0.77 \\
    $c_{2}(0, 400)$ & 0.23 & 0.23 & 0.22 \\
    $c_{2}(20, 400)$ & 0.055 & 0.056 & 0.054 \\
    $a(0, 400)$ & 0.20 & 0.21 & 0.20 \\
    $a(20, 400)$ & 0.84 & 0.84 & 0.83 \\
    $g(0, 400)$ & 3.23 & 3.23 & 3.24 \\
    $g(20, 400)$ & 0 & 0 & 0 \\
    $[gc_{1}](0, 400)$ & 1.04 & 1.03 & 1.04 \\
    $[gc_{1}](20, 400)$ & 0 & 0 & 0 \\
    $[gc_{2}](0, 400)$ & 0.73 & 0.74 & 0.72 \\
    $[gc_{2}](20, 400)$ & 0 & 0 & 0 \\
  \end{tabular}
  \end{ruledtabular}
\end{table}

\begin{figure*}
    \centering
    \includegraphics[scale=0.35]{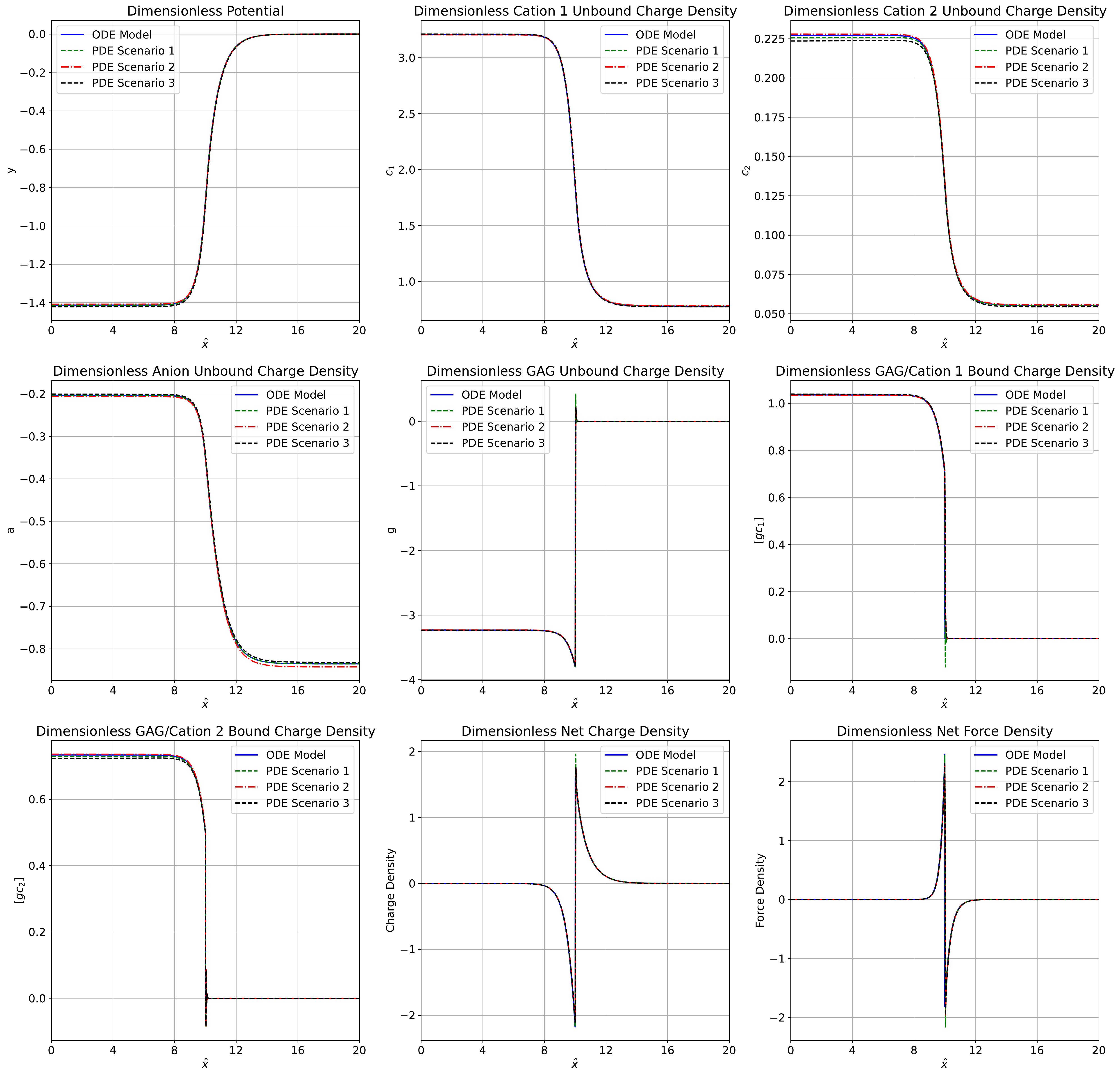}
    \caption{Overlay of numerical results from ODE model and 3 scenarios performed
    for the time-dependent model.}
    \label{ode_pde_res}
\end{figure*}

Using the results in Table \ref{pde_result}, we derived the necessary input parameters for our ODE
model in Table \ref{ode_params_from_pde}. The results from our ODE model are overlaid with the
results of the 3 scenarios from our time-dependent model in Figure \ref{ode_pde_res}.

\begin{table}
  \caption{ODE Mode Input Parameters Derived from Time-Dependent Model Steady State Solution.}
  \label{ode_params_from_pde}
  \begin{ruledtabular}
  \begin{tabular}{ll}
    \textbf{Parameters}                       & \textbf{Volume Model}   \\
    \hline
    $\hat{\ell}$ & 10 \\
    $\hat{L}$ & 20 \\
    $\widetilde{K}_{1}$ & 10 \\
    $\widetilde{c}_{1}$ & 0.78 \\
    $\widetilde{K}_{2}$ & 1 \\
    $\widetilde{c}_{2}$ & 0.055 \\
    $\bar{g}$ & 5 \\
    $\hat{\sigma}_{1}$ & 0 \\
    $\hat{\sigma}_{2}$ & 0 \\
  \end{tabular}
  \end{ruledtabular}
\end{table}

There is excellent agreement in the steady state solution between our ODE model and the 3
scenarios performed with the time-dependent model. There is a slight artifact that can be seen
in the GAG related charge density curves at $\hat{x}=10$ in the time-dependent model solutions
that is due to the step function used for the boundary between the GAG and salt regions. This
is strictly numerical in nature and can be resolved by using a smoothed step function. Figure
\ref{ode_pde_res} also includes a plot of the dimensionless net charge density, which is the
right-hand side of equation (\ref{pde_pot}) at steady state. Using this, we can define a
dimensionless net Coulombic force density as
\beq
f = -(c_{1}+c_{2}-a-g)\frac{\partial y}{\partial x},
\eeq
which is the dimensionless net charge density multiplied by the spatial derivative of the
dimensionless potential. We see that both the net charge density and the net force density
are zero everywhere except in a region near the boundary between the GAG and salt regions
at $\hat{x}=10$. From the force density, we see that there is a \textit{pinching} effect at the
boundary where the force in the GAG region at the boundary is towards the right (salt region)
and the force in the salt region at the boundary is towards the left (GAG region). Such a force calculation is more complicated if there is an overshoot in potential as seen in Figures \ref{hy_NaCl_fig} and \ref{he_NaCl_fig} for GAG brushes with a sodium cation. In these cases, the Coulombic force traversing the brush edge will be right-left-right-left and Born hydration forces also need to be considered in the force balance. 

\section{Conclusion}

We have developed steady-state and transient mean-field models for GAG brushes that capture ion-partitioning, ion-pairing, and dielectric decrement effects in GAG brushes. Our models build upon previous Poisson-Boltzmann models \cite{dean2003molecular, sterling2017electro}
and generate the same predicted performance when applying our model in the appropriate limits.
We have also shown that our model predictions agree well with the molecular simulation
data \cite{sterling2021ion}, however our model requires the anion Born radius to
be much smaller than typical values found in the literature \cite{sun2020analysis}.
The requirement that this unrealistic, nonphysical ion hydration radius be used
to match experimental or molecular dynamics simulation data appears to be a known
challenge for Poisson-Boltzmann based models \cite{eakins2021modeling}. Furthermore, there is evidence from neutron diffraction experiments on concentrated salt solutions that the chloride ion disrupts the hydrogen-bond network of bulk water much less than the cations; effects that are not captured using the Born hydration formulation implemented  herein \cite{mancinelli2007hydration}. Another explanation is that there are additional energy terms such as GAG dipole effects that our model is neglecting. Further research
is needed in this area.

Our time-dependent PNP models are shown to be stable and to relax to the PB models, meaning that the imposed jump conditions result in interface profiles that occur at steady-state. Deviations from the steady state profiles that conserve species are shown to relax to the appropriate steady-state.

While we have presented a model where the permittivity is a simple function of $x$,
others have proposed a dependence of the permittivity on ion concentrations
\cite{gavish2016dependence, lopez2018multiionic}. Future work will entail
exploring the appropriate relationship between permittivity and ion concentration and
incorporating this type of relationship into our model. We will also continue the
development of our time-dependent model by incorporating time varying permittivity, the
Born energy, and additional energy terms as mentioned above.

The models presented here offer biophysical detail of biological environments that have anionic beds of macromolecules that attain electroneutrality via atomic cations and cationic residues of proteins. The extent to which such biophysical detail can be used to develop new diagnostics, drug-delivery platforms, or therapeutics remains to be seen.

%\begin{acknowledgments}
%We wish to acknowledge the support of the author community in using
%REV\TeX{}, offering suggestions and encouragement, testing new versions,
%\end{acknowledgments}

%\nocite{*}

\bibliography{ref}% Produces the bibliography via BibTeX.

\end{document}